\documentclass[letterpaper]{JHEP3}

\usepackage{amsmath}
\usepackage{epsfig,multicol,bbm}
\usepackage{graphicx}
\usepackage{bm}
\usepackage{epsfig}
\usepackage{url}

\renewcommand{\thesection}{\arabic{section}}
\renewcommand{\emph}[1]{{\it #1}}

\newcommand{\be}{\begin{equation}}
\newcommand{\ee}{\end{equation}}
\newcommand{\nn}{\nonumber}

\newcommand{\GeV}{\text{ GeV}}
\newcommand{\TeV}{\text{ TeV}}

\newcommand{\Eq}[1]{Eq.~\eqref{#1}}
\newcommand{\Tab}[1]{Table~\ref{#1}}
\newcommand{\Sec}[1]{Sec.~\ref{#1}}
\newcommand{\Fig}[1]{Fig.~\ref{#1}}
\newcommand{\App}[1]{App.~\ref{#1}}
\newcommand{\Ref}[1]{Ref.~\cite{#1}}

\title{Semi-annihilation of Dark Matter}

\author{Francesco D'Eramo and Jesse Thaler \\
Center for Theoretical Physics, Massachusetts Institute of Technology,
Cambridge, MA 02139, USA \vspace{0.05in} \\
E-mail: \email{fderamo@mit.edu}, \email{jthaler@jthaler.net} }
 
\preprint{MIT-CTP 4136}

\abstract{We show that the thermal relic abundance of dark matter can be affected by a new type of reaction:  semi-annihilation.  Semi-annihilation takes the schematic form $\psi_i \psi_j \rightarrow \psi_k \phi$, where $\psi_i$ are stable dark matter particles and $\phi$ is an unstable state.  Such reactions are generically present when dark matter is composed of more than one species with ``flavor'' and/or ``baryon'' symmetries.  We give a complete set of coupled Boltzmann equations in the presence of semi-annihilations, and study two toy models featuring this process.   Semi-annihilation leads to non-trivial dark matter dynamics in the early universe, often dominating over ordinary annihilation in determining the relic abundance.  This process also has important implications for indirect detection experiments, by enriching the final state spectrum from dark matter (semi-)annihilation in the Milky Way.}

\begin{document}

\section{Introduction}
\label{sec:intro}

The existence of dark matter is one of the best motivations for physics beyond the standard model (SM).   Evidence for dark matter has accumulated at vastly different length scales---from galactic scales and clusters of galaxies to global scales of hundreds of megaparsecs \cite{Jungman:1995df,Bergstrom:2000pn,Bertone:2004pz}. However, all of these observations infer the existence of dark matter through its gravitational effects alone. In particular, they do not tell us the nature, origin, or composition of this important component of our universe, which is not explained by any SM degree of freedom.

A particularly well-motivated class of dark matter candidates are so-called Weakly Interacting Massive Particles, or WIMPs, whose abundance is determined through thermal freeze-out.  In the Lee-Weinberg scenario \cite{Lee:1977ua}, WIMPs $\psi$ thermalize in the early universe through the annihilation reaction $\psi \bar{\psi} \rightarrow \phi \phi'$, where $\phi$ and $\phi'$ are SM degrees of freedom, until their interactions freeze out.  A standard relic abundance calculation \cite{Krauss:1983ik,Scherrer:1985zt,Kolb:1985nn,Srednicki:1988ce,Gondolo:1990dk,Kolb:1990vq,Bottino:1993zx} shows that the dark matter mass density today depends only logarithmically on the relic particle mass and scales inversely with the total annihilation cross section at freeze-out, $\Omega_{{\rm DM}} h^2 \propto \langle \sigma v\rangle^{-1}$. It is remarkable that for a dark matter mass between $10 \GeV-10 \TeV$ and an electroweak annihilation cross section, one gets a relic abundance in the ballpark to account for dark matter.   This fact is often referred to as the ``WIMP miracle'', and in the context of the gauge hierarchy problem, it is suggestive that the same particles one might introduce to stabilize the Fermi scale could also explain the dark matter in the universe.  

Since the WIMP paradigm is one of the best explanations for dark matter and since thermal freeze-out is so predictive, it is important to know how to correctly compute the dark matter thermal relic density.  This is particularly important for models where the relic computation cannot be reduced to the Lee-Weinberg scenario.  As we will argue, the thermal abundance of dark matter can be dramatically affected by the presence of a new dark matter interaction, which we call ``semi-annihilation''.  

Semi-annihilation occurs when dark matter is stabilized by a larger symmetry than just $Z_2$.  In the simplest case with just one dark matter species $\psi$, there can be an additional allowed reaction
\begin{equation}
\psi \psi \rightarrow \psi \phi
\end{equation}
which preserves a  $Z_3$ symmetry.  Here, $\phi$ is a SM state or a new particle which decays to the SM.  We see that unlike ordinary annihilation where the total dark matter number changes by two units, in semi-annihilation the total dark matter number changes by only one unit.  More generally, dark matter can be composed of more than one stable component $\psi_i$, and these relic particles can have non-trivial mutual interactions.  In this case, a more general semi-annihilation reaction is possible,
\begin{equation}
\psi_i \psi_j \rightarrow \psi_k \phi ,
\end{equation}
which often occurs if dark matter is stabilized by  ``baryon'' and/or ``flavor'' symmetries, as in QCD-like theories. Such reactions are also allowed in models where dark matter is composed of non-Abelian gauge bosons~\cite{Hambye:2008bq,Hambye:2009fg,Arina:2009uq}. In this paper, we study explicit examples of such models and find that to correctly compute the relic abundance, semi-annihilation must be included. In fact, semi-annihilation can dominate over standard annihilation for some regions of parameter space. 

To understand how semi-annihilation fits into the WIMP paradigm, consider a more general framework for dark matter interactions.  Start by assuming the existence of a new dark sector composed of $N$ particles $\psi_i$ which can be either stable or unstable.  The possible reactions involving $\psi_i$ which can take place in the early universe are
\begin{equation}
a) \,  \psi_i \psi_j \rightarrow \phi \phi^{\prime},\;\;\; b) \,  \psi_i \phi \rightarrow \psi_j \phi^{\prime},\;\;\; c) \,  \psi_i \psi_j \rightarrow \psi_k \phi,\;\;\; d) \, \psi_i \psi_j \rightarrow \psi_k \psi_m,\;\;\; e) \, \psi_i \rightarrow \psi_j \phi,
\label{eq:abcde}
\end{equation}
where again $\phi$ and $\phi^{\prime}$ are light degrees of freedom in thermal equilibrium with the SM.   Different dark matter thermal freeze-out scenarios depend on which of the reactions $a)$--$e)$ are active, and we summarize these main possibilities in \Tab{tab:reactions}.

\TABLE[t]{
\begin{tabular}{c|c|c|c|c|}
& $\psi_i \psi_j \rightarrow \phi \phi^{\prime}$ & $\psi_i \phi \rightarrow \psi_j \phi^{\prime}$ & $\psi_i \psi_j \rightarrow \psi_k \phi$ & $\psi_i \rightarrow \psi_j \phi$ \\ 
\hline 
Lee-Weinberg & $\checkmark(i=j)$ & $\checkmark(i=j)$ & $\times$ & $\times$ \\
Co-annihilation & $\checkmark$ & $\checkmark$ & $\times$ & $\checkmark$\\
Multi-component & $\checkmark(i=j)$ & $\checkmark(i=j)$ & $\times$ & $\times$\\
Semi-annihilation & $\checkmark(i=j)$ & $\checkmark(i=j)$ & $\checkmark$ & $\times$ \\
\hline
\end{tabular}
\label{tab:reactions}
\caption{Different dark matter freeze-out scenarios and allowed reactions.  We do not include the reaction $\psi_i \psi_j \rightarrow \psi_k \psi_m$ in this table since it is always present, and we explicitly indicate when only the diagonal $(i= j)$ contribution is allowed.}
}

\begin{description}
\item{\bf Lee-Weinberg}:
The simplest case is the Lee-Weinberg scenario \cite{Lee:1977ua}, where $N=1$ and the only allowed reactions are $a)$, $b)$, and $d)$, with $i=j$.  Chemical freeze-out is determined by the annihilation reaction $a)$, and kinetic freeze-out is determined by $b)$.  Reaction $d)$ plays no role in the thermal relic computation.
\item{\bf Co-annihilation}:
A slight variation of the standard case is when $N>1$ but there is only one stable dark matter species.  Here the relevant reactions are $a)$, $b)$, $d)$, and $e)$.  This is the case in the minimal supersymmetric standard model with $R$-parity, where all heavier particles are unstable and eventually decay to the lightest one via reaction $e)$.  In principle, this co-annihilation case involves a system of $N$ coupled Boltzmann equations, but as we will review, as long as reactions of type $b)$ with $i \not= j$ are effective at freeze-out, it is possible to compute the relic density via standard methods \cite{Griest:1990kh}.
\item{\bf Decoupled Multi-Component}:
These first two examples assume dark matter to be composed of a single particle, but more generally, dark matter could be composed of more than one stable component.  Many such multi-component dark matter models have been proposed (see e.g.\ \cite{Boehm:2003ha,Ma:2006uv,Hur:2007ur,Hur:2007ur,Feng:2008ya,Fairbairn:2008fb,SungCheon:2008ts,Zurek:2008qg,Morrissey:2009ur}). The standard approach in multi-component models is to assume that each particle thermalizes independently of the others, thus the total dark matter density today is $\Omega_{\rm DM} = \sum_i \Omega_i$ where the sum runs over all the thermal relics.  Said in the language of \Eq{eq:abcde}, only the diagonal ($i=j$) reactions of type $a)$ and $b)$ are present, and reactions $c)$ and $e)$ are forbidden.  Reaction $d)$ may or may not be present in such models, and is relevant for calculating the relic density \cite{SungCheon:2008ts}.

\item{\bf Semi-annihilation}:
The focus of this paper is on the reaction of type $c)$, which to our knowledge first appeared in \Ref{Hambye:2008bq,Hambye:2009fg,Arina:2009uq}.\footnote{In the models considered in \Ref{Hambye:2008bq,Hambye:2009fg,Arina:2009uq}, a custodial symmetry effectively reduced the relic abundance computation to a single particle system.   Here, we consider semi-annihilation in multi-component dark matter models where there is no such simplification.}  
As long as the triangle inequality $m_k < m_i+m_j$ is satisfied (as well as its crossed versions), the semi-annihilation reaction $\psi_i \psi_j \rightarrow \psi_k \phi$ can take place without making any relic particle unstable.  Semi-annihilation implies that the relic particles have non-trivial mutual interactions, and typically both reactions of type $c)$ and $d)$ are important in determining the dark matter relic abundance.  For simplicity, we focus on the case where each particle in the dark sector is absolutely stable, in which case reaction $e)$ is forbidden.  In addition, only the diagonal reactions of type $a)$ and $b)$ are allowed, since off-diagonal contributions would imply that the heavier particle would be unstable by crossing symmetry. As we will see, the absence of off-diagonal reactions of type $b)$ makes these models more difficult to study than the standard cases.
\end{description}

The remainder of this paper is structured as follows.  In \Sec{sec:Z3} we present a toy model where dark matter is composed of a single component but semi-annihilations are present and can dominate dark matter production in the early universe.   In \Sec{sec:semimulti} we introduce example multi-component models where semi-annihilation reactions are present, and we give a complete set of Boltzmann equations which can be solved to obtain the relic density today.  In particular, we argue that a semi-analytical solution analogous to the Lee-Weinberg scenario or co-annihilation is not in general possible, and thus the equations have to be solved numerically.  We present a minimal multi-component toy model  with semi-annihilations in \Sec{sec:bbchi} and numerically study the effects of semi-annihilation on the relic abundance.  We explore the effects of semi-annihilations on indirect detection experiments in \Sec{sec:Indirect}, and conclude in \Sec{sec:Conclusions}.  Computational details, as well as an example supersymmetric QCD model with $N_f = N_c+1$, are given in the appendices.

\section{Semi-annihilation with a $Z_3$ symmetry}
\label{sec:Z3}

To introduce semi-annihilation, we present a simple case where semi-annihilations play a significant role in the thermal relic calculation.  We assume dark matter to be composed of one single stable particle $\chi$, a complex scalar with mass $m_{\chi}$, which is stabilized by a $Z_3$ symmetry.\footnote{For a concrete realization of dark matter models with a $Z_3$ symmetry and with semi-annihilation playing a subdominant role see \Ref{Agashe:2004ci,Agashe:2004bm}.  See also \Ref{Ma:2007gq} and \Ref{Walker:2009ei,Walker:2009en}.} The $\chi$ particle interacts with a real scalar $\phi$, which eventually decays to SM states. The interaction Lagrangian is
\begin{equation}
\mathcal{L}_{Z_3} = m_{\chi}^2 \, \chi^\dag \chi + a_1\, \chi^\dag \chi \phi + a_2 \, \chi^\dag \chi \phi^2 + a_3 \, \chi\chi\chi\phi - V(\phi) , 
\end{equation}
where $V(\phi)$ contains additional interactions for $\phi$ alone.  This function contains both a scalar potential for $\phi$ and couplings with SM fields, 
\begin{equation}
V(\phi) = c_1 \, \phi^3 + c_2 \, \phi ^4 + d_1 \, \phi B_{\mu\nu} B^{\mu\nu} + d_2 \, \phi H^{\dag} H + \cdots,
\label{eq:Vphi}
\end{equation}
where $B_{\mu\nu}$ is the hypercharge field strength, $H$ is the Higgs doublet and the $\cdots$ stand for other possible interactions. The model and its symmetries are summarized in \Tab{tab:Z3}.

\TABLE[t] {
\parbox{5in}{ 
\begin{center}
\begin{tabular}{|c|c|c|}
\hline
 & spin & $Z_3$ \\ 
\hline\hline
$\chi$ & complex scalar & $\left(-1\right)^{1/3}$ \\
$\phi$ & real scalar & $0$\\
\hline
\end{tabular}
\end{center}
}
\caption{Field content and symmetries of the model with a $Z_3$ symmetry.}
\label{tab:Z3}
}

As long as the coupling strength of $\phi$ to SM fields is sufficiently large, $\phi$ will be kept in thermal equilibrium with the SM.  Then, the $\chi$ particles are kept in thermal equilibrium in the early universe by the reactions $\chi \bar{\chi} \rightarrow \phi \phi$ and $\chi \chi \rightarrow \bar{\chi} \phi$, as shown in \Fig{fig:Z3FeynDiag}. While we focus on the case of a field $\phi$ ``portal'' to the SM  \cite{Finkbeiner:2007kk,Pospelov:2007mp,ArkaniHamed:2008qn,Nomura:2008ru}, in more general dark matter scenarios $\phi$ could be a SM field itself \cite{Burgess:2000yq,Patt:2006fw,MarchRussell:2008yu}.

\FIGURE[t]{
$\begin{array}{c@{\hspace{1in}}c}
\includegraphics[scale=0.5]{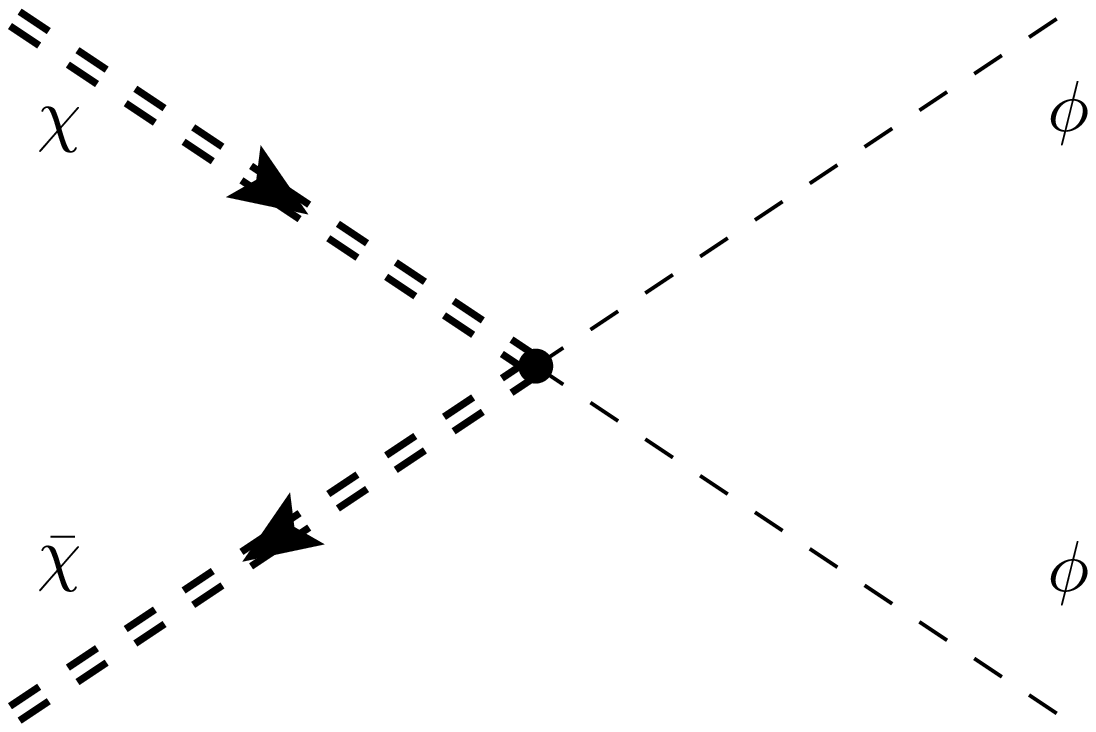} &
\includegraphics[scale=0.5]{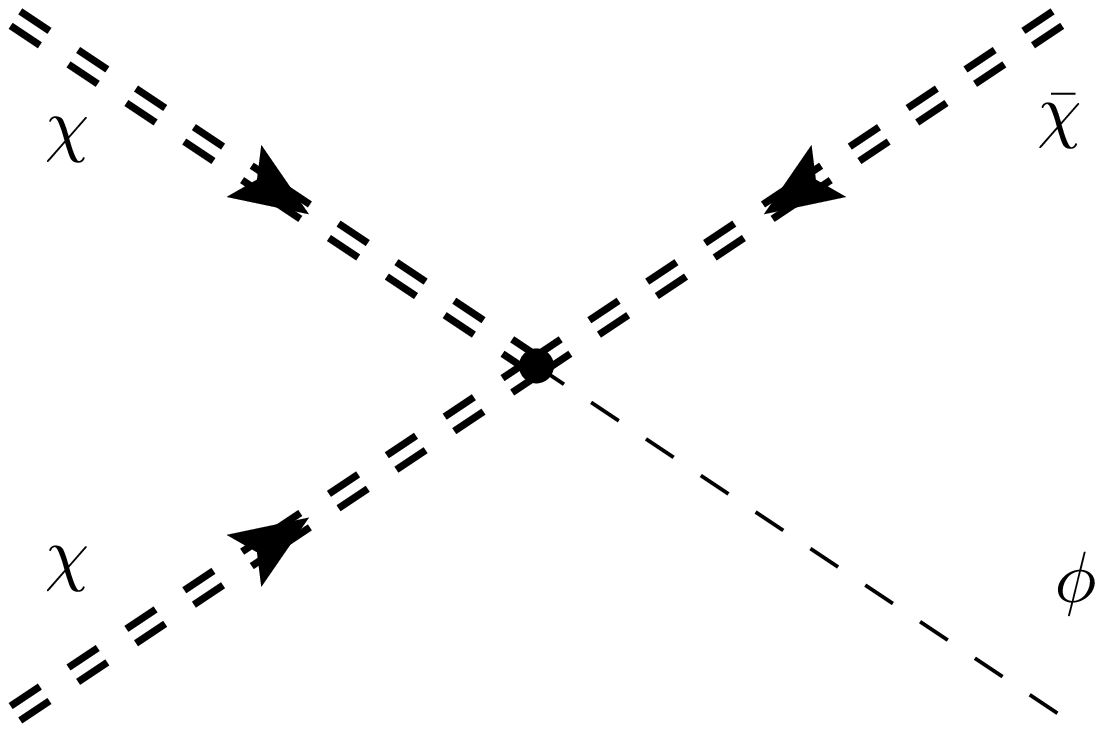} \\ [0.1cm]
\mbox{\bf (a)} & \mbox{\bf (b)}
\end{array}$
\caption{The reactions that keep $\chi$ in thermal equilibrium in the model with $Z_3$ symmetry: ({\rm a}) annihilation; ({\rm b}) semi-annihilation. The $\chi$ fields are drawn as double lines.  The field $\phi$ decays to SM states such that the whole system is in thermal contact with the SM before freeze-out.}
\label{fig:Z3FeynDiag}
}

\subsection{Boltzmann equation}
The Boltzmann equation describing the number density evolution of $\chi$ is\footnote{For a brief derivation of the Boltzmann equation in an expanding universe, see \App{app:Boltzmann}.  The following Boltzmann equation also appears in \Ref{Hambye:2009fg,Arina:2009uq}, despite the fact that they have a multi-component model.   In that case, a custodial symmetry implies a degeneracy for the dark matter species number densities, reducing the relic abundance calculation to a single differential equation.}
\begin{equation}
\frac{d n_{\chi}}{d t} + 3 H n_{\chi} = - \langle \sigma v \rangle_{\chi\bar{\chi}\rightarrow\phi\phi} \left[n_{\chi}^2 - n_{\chi}^{\text{eq}\, 2}\right] 
- \langle \sigma v \rangle_{\chi\chi\rightarrow\bar{\chi}\phi} \left[n_{\chi}^2 - n_{\chi} n_{\chi}^{\text{eq}}\right],
\label{eq:z3boltzmann}
\end{equation}
where $H$ is the Hubble parameter and $n_{\chi}^{\text{eq}}$ is the equilibrium number density distribution.  In the $\langle \sigma v \rangle_{\chi\chi\rightarrow\bar{\chi}\phi}=0$ limit we recover the familiar Lee-Weinberg scenario.   To get an understanding for the behavior of this Boltzmann equation, we restrict ourselves to $s$-wave (semi-)annihilation, and we choose as free parameters the $s$-wave amplitudes for the two processes, $\eta_{a}$ and $\eta_{s}$ respectively. The thermal averages of the $s$-wave cross sections can be obtained using \Eq{eq:easythermal} from \App{app:thaverages}:\footnote{The factor of $3/4$ comes from the phase space suppression in semi-annihilation.}
\begin{equation}
\langle \sigma v \rangle_{\chi\bar{\chi}\rightarrow\phi\phi} = \frac{\eta^2_{a}}{32\pi m_{\chi}^2}, \qquad \langle \sigma v \rangle_{\chi\chi\rightarrow\bar{\chi}\phi} = \frac{3}{4} \frac{\eta^2_{s}}{32\pi m_{\chi}^2}.
\end{equation}

\subsection{Semi-analytical solution}

The above Boltzmann equation can be solved semi-analytically in analogy with the Lee-Weinberg calculation.  We rewrite \Eq{eq:z3boltzmann} by introducing a dimensionless time variable $x=m_{\chi}/T$ and comoving number density $Y_{\chi}=n_{\chi}/s$, where $s$ is the entropy density of the relativistic degrees of freedom. The Boltzmann equation in these new variables is
\begin{equation}
\frac{d Y_{\chi}}{d x} = -\frac{\lambda_a}{x^2}\left[Y_{\chi}^2 - Y_{\chi}^{\text{eq}\, 2}\right] -\frac{\lambda_s}{x^2}\left[Y_{\chi}^2 - Y_{\chi} Y_{\chi}^{\text{eq}}\right], \qquad \lambda_i \equiv \frac{s(T=m_{\chi})}{H(T=m_{\chi})}\langle\sigma v \rangle_i .
\label{eq:z3boltzmannVar}
\end{equation}
We can give a semi-analytical solution to this equation following the method in \Ref{Kolb:1990vq}, by solving the equation in two regimes, early and late times, and then matching the two solutions at the freeze-out point.  For convenience, we define the function $\Delta = Y_{\chi} - Y_{\chi}^{\text{eq}}$. 

At early times, the number density is very well approximated by its equilibrium value, thus we impose $d \Delta / dx=0$ in the Boltzmann equation.  The use of the Maxwell-Boltzmann equilibrium distribution for $Y_{\chi}^{\text{eq}}$ is well-justified for a cold relic.  Eventually a point denoted ``freeze-out'' is reached when the interaction rate is not fast enough compared with the expansion rate to maintain thermal equilibrium.  The freeze-out point $x_f$ is defined as $\Delta (x_f) = c\, Y^{\text{eq}}_{\chi}(x_f)$, where $c$ is a numerical constant of the order $1$.  Since we are making the $s$-wave approximation, we take $c=\sqrt{2}-1$ as suggested by \Ref{Kolb:1990vq}.  After freeze-out, the equilibrium distribution is exponentially suppressed and we can safely neglect it in the Boltzmann equation and solve directly for $Y_\chi$.  We finally match the two solutions at the freeze-out point $x_f$. 

The freeze-out value $x_f$ is found by solving the equation
\begin{equation}
x_f = \log\left[0.038\,c(c+2)\langle \sigma v \rangle_{\chi\bar{\chi}\rightarrow\phi\phi} \frac{g_{\chi}\,m_{\chi}\,M_{\text{Pl}}}{\sqrt{g_{*}\,x_f}}\right] + 
\log\left[1 + \frac{c+1}{c+2}\,\frac{\langle \sigma v \rangle_{\chi\chi\rightarrow\bar{\chi}\phi}}{\langle \sigma v \rangle_{\chi\bar{\chi}\rightarrow\phi\phi}}\right]
\label{eq:xfZ3},
\end{equation}
where $g_{\chi}=1$ for a scalar field, $g_{*}$ is the effective number of relativistic degrees of freedom at the time of the freeze-out, and the Planck scale is $M_{\text{Pl}} = 1.22 \times 10^{19}\,{\rm GeV}$.  The first term on the right-hand side of \Eq{eq:xfZ3} is the solution we usually have in the Lee-Weinberg scenario.  Thus we see that the effect of semi-annihilation is to shift the freeze-out temperature by only a small logarithmic amount. 

The mass density of the $\chi$ particle today is given by $\rho_{\chi,\,0} = m_{\chi} \,s_0\,Y_{\chi}(\infty)$, and as usual can be expressed as the fraction of the critical energy density
\begin{equation}
\Omega_{\chi} h^2 = 2 \times \frac{1.07 \times 10^9 \, {\rm GeV^{-1}}}{\sqrt{g_*} M_{\text{Pl}} J(x_f)},\qquad\qquad J(x_f) = \int_{x_f}^{\infty} dx\,\frac{\langle \sigma v \rangle_{\chi\bar{\chi}\rightarrow\phi\phi} + \langle \sigma v \rangle_{\chi\chi\rightarrow\bar{\chi}\phi}}{x^2} ,
\label{eq:Omegah2Z3}
\end{equation}
where the $2$ factor takes into account the antiparticle degree of freedom and the annihilation integral includes both annihilations and semi-annihilations. 

\subsection{Results}
The results of solving \Eq{eq:z3boltzmannVar} are shown in \Fig{fig:numvsan} for the mass value $m_{\chi} =  1\, {\rm TeV}$.  We consider three representative cases: annihilation only, semi-annihilation only, and both processes present with equal amplitudes. In each case the parameters are chosen in order to get the observed relic abundance, $\Omega_{\chi} h^2 \simeq 0.1$ \cite{Dunkley:2008ie}. We plot both the numerical and the semi-analytical solutions, and we find a good agreement between them.  In \Fig{fig:numvsan}(d), we show the ratios between the various cases, where we see that as we increase the semi-annihilation contribution, it takes longer for the system to reach the freeze-out density.  This behavior is expected from \Eq{eq:z3boltzmannVar}, since the collision operator for semi-annihilation has a positive, linear dependence on $Y_\chi$ that somewhat counteracts the overall depletion of $\chi$ particles.

\FIGURE[t]{
$\begin{array}{cc}
\includegraphics[scale=0.63]{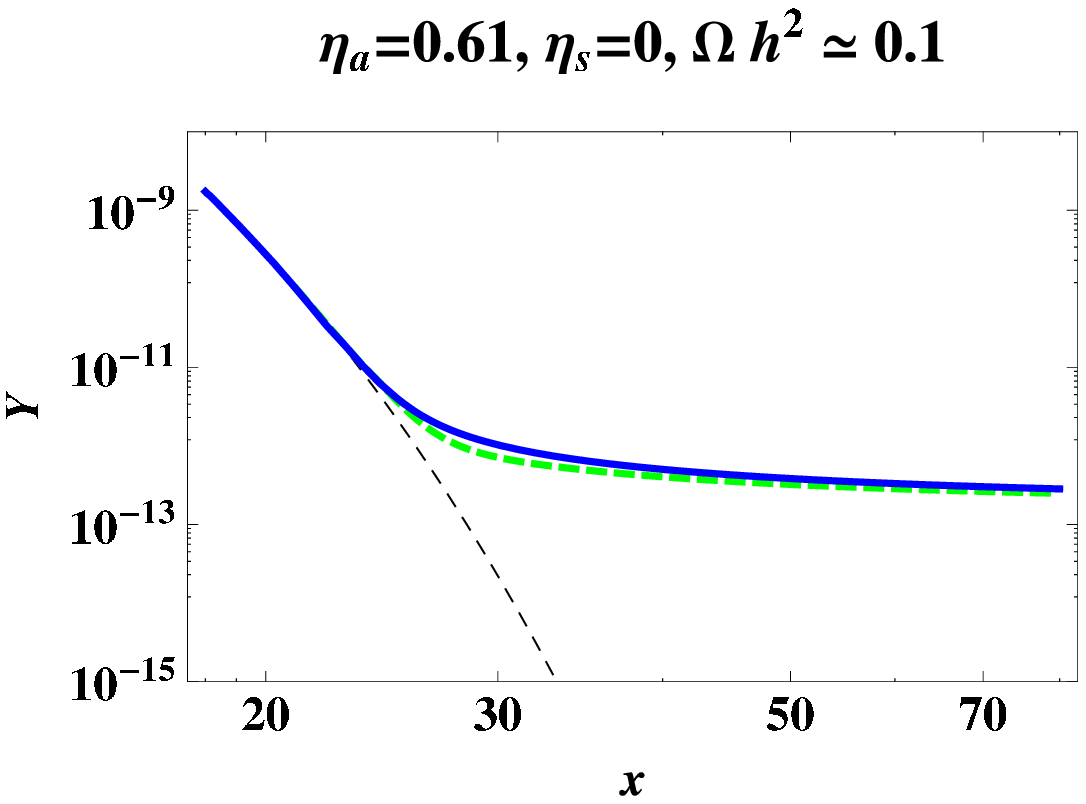} & \hspace{0.4cm} \includegraphics[scale=0.63]{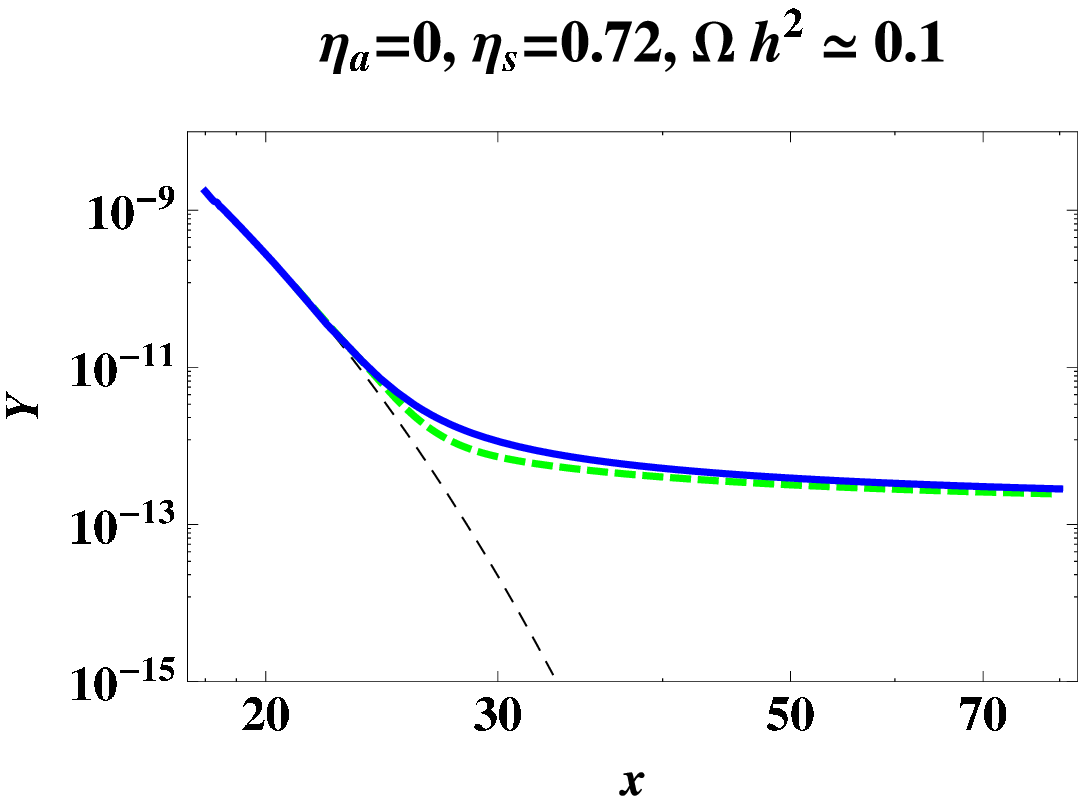} \\ 
\mbox{\bf (a)} & \mbox{\bf (b)} \\[0.3cm]
\includegraphics[scale=0.63]{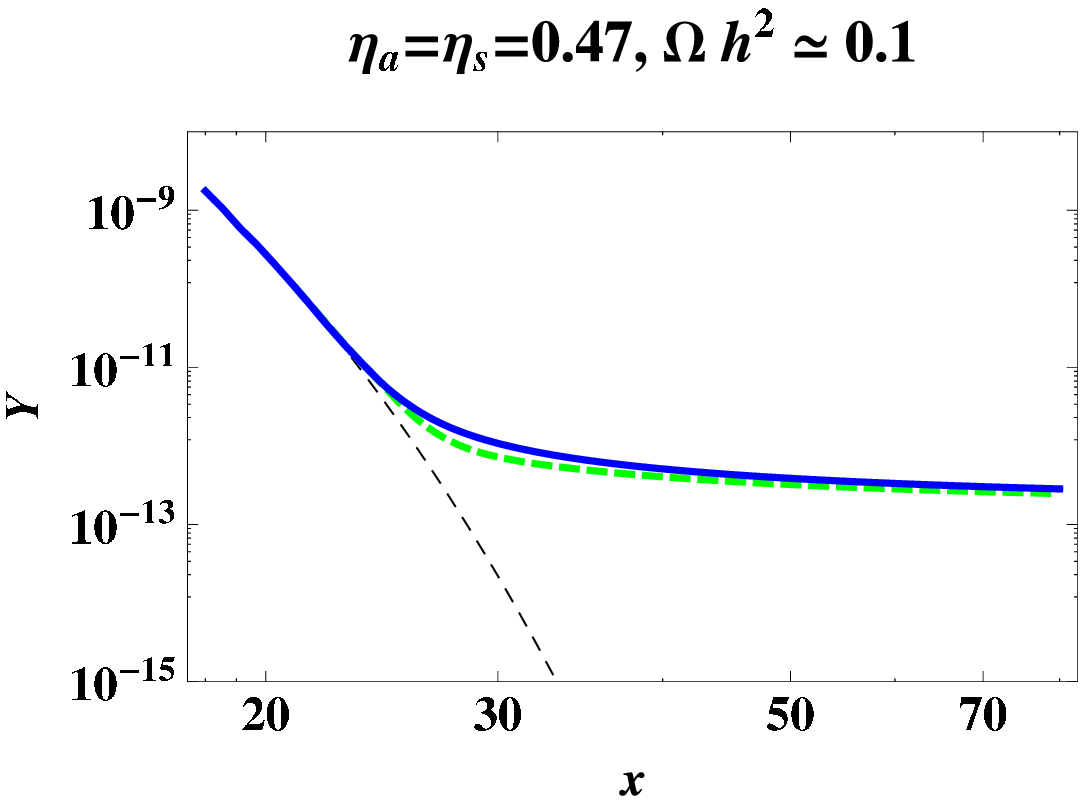} & \hspace{0.4cm} \includegraphics[scale=0.68]{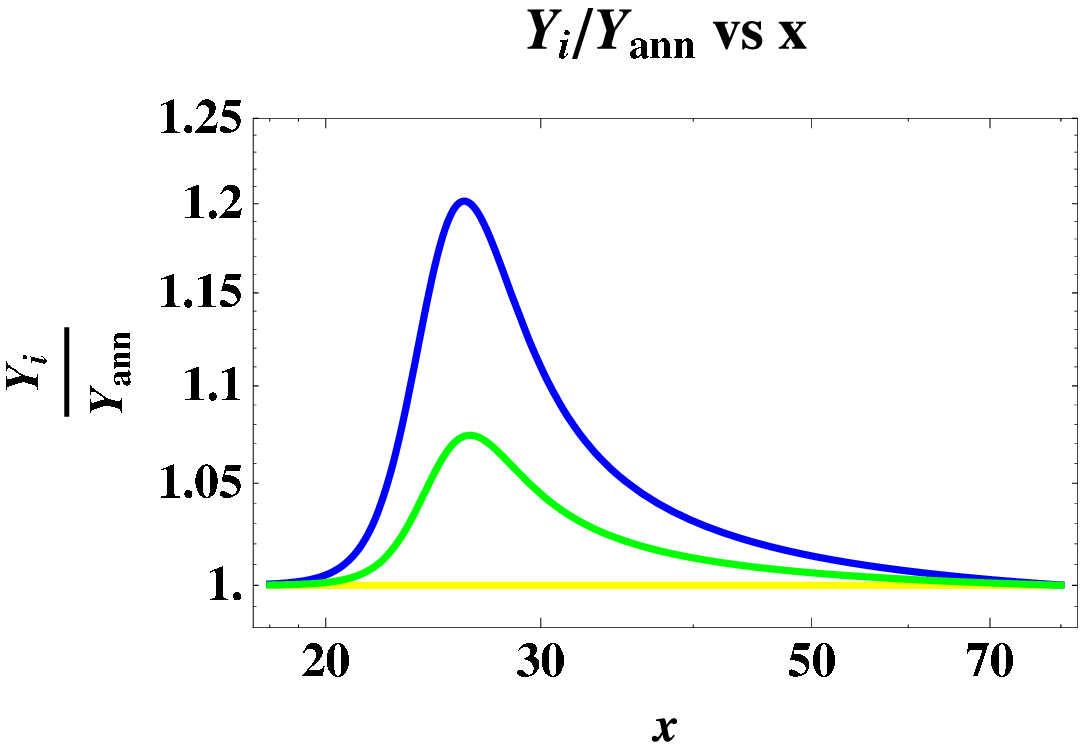} \\
\mbox{\bf (c)} & \mbox{\bf (d)} \\
\end{array}$
\caption{Comoving number density $Y(x)$ in the $\chi$ model with a $Z_3$ symmetry for $m_{\chi} =  1\, {\rm TeV}$.  Three cases are considered: ({\rm a}) pure annihilation, ({\rm b}) pure semi-annihilation, ({\rm c}) equal amplitudes for the two processes. In all the three cases, the parameters $\eta_a$ and $\eta_s$ are chosen in order to get the WMAP value for the relic density today, $\Omega_{{\rm DM}} h^2 \simeq 0.1$. We plot the equilibrium solution (dashed black), the numerical solution (solid blue) and the semi-analytical solution (dashed green).  To emphasize the subtle differences between the cases, in ({\rm d}) we show the ratio between each numerical solution and the one for pure annihilation, where pure annihilation is yellow, pure semi-annihilation is blue, and the mixed case is green.}
\label{fig:numvsan}
}

In \Fig{fig:chichichiphi}, we identify the region of the $\left(\eta_a, \eta_s\right)$ plane which gives the observed dark matter relic density in our universe.  We consider two different dark matter particle masses, $m_{\chi} = 1\, {\rm TeV}$ and $m_{\chi} = 5 \, {\rm TeV}$ respectively, and we shade the region where the dark matter relic density is within the WMAP $95\%$ CL region \cite{Dunkley:2008ie}, namely $0.0975 \leq \Omega_{{\rm DM}} h^2 \leq 0.1223$. In single component dark matter models, the conventional assumption is that we have only annihilation (namely $\eta_s = 0$), thus once we fix the dark matter mass $m_{\chi}$ the amplitude $\eta_a$ is uniquely determined by the relic density measurement. However, if we allow also semi-annihilations, a wider region of parameter space gives the correct relic density.  Such a region corresponds roughly to $4 \eta_a^2 + 3 \eta_s^2 \simeq 1.5$, thus the thermal production of the relic particles in the early universe can also be completely controlled by semi-annihilations.

\FIGURE[t]{
$\begin{array}{cc}
\includegraphics[scale=0.66]{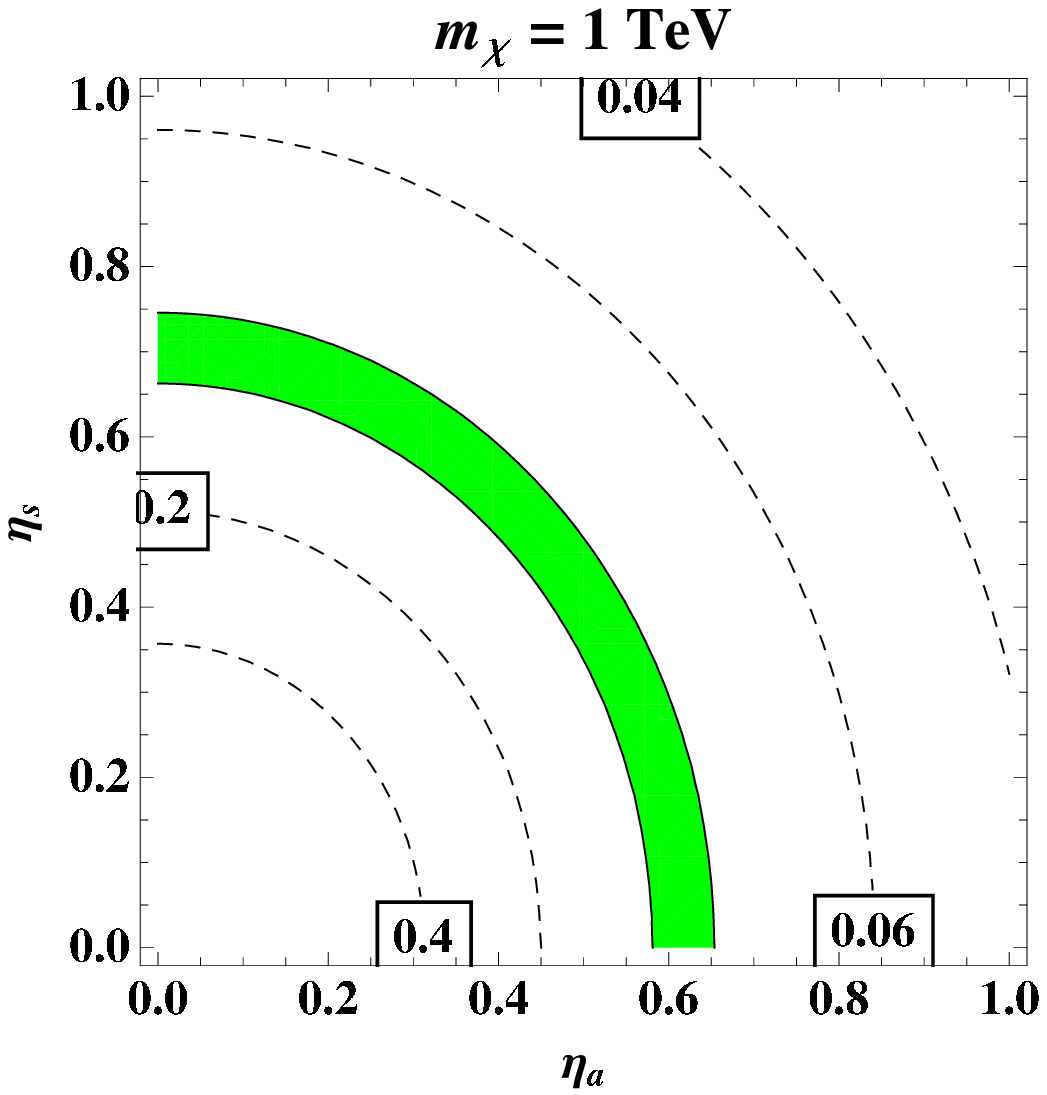} &\hspace{0.2cm}
\includegraphics[scale=0.63]{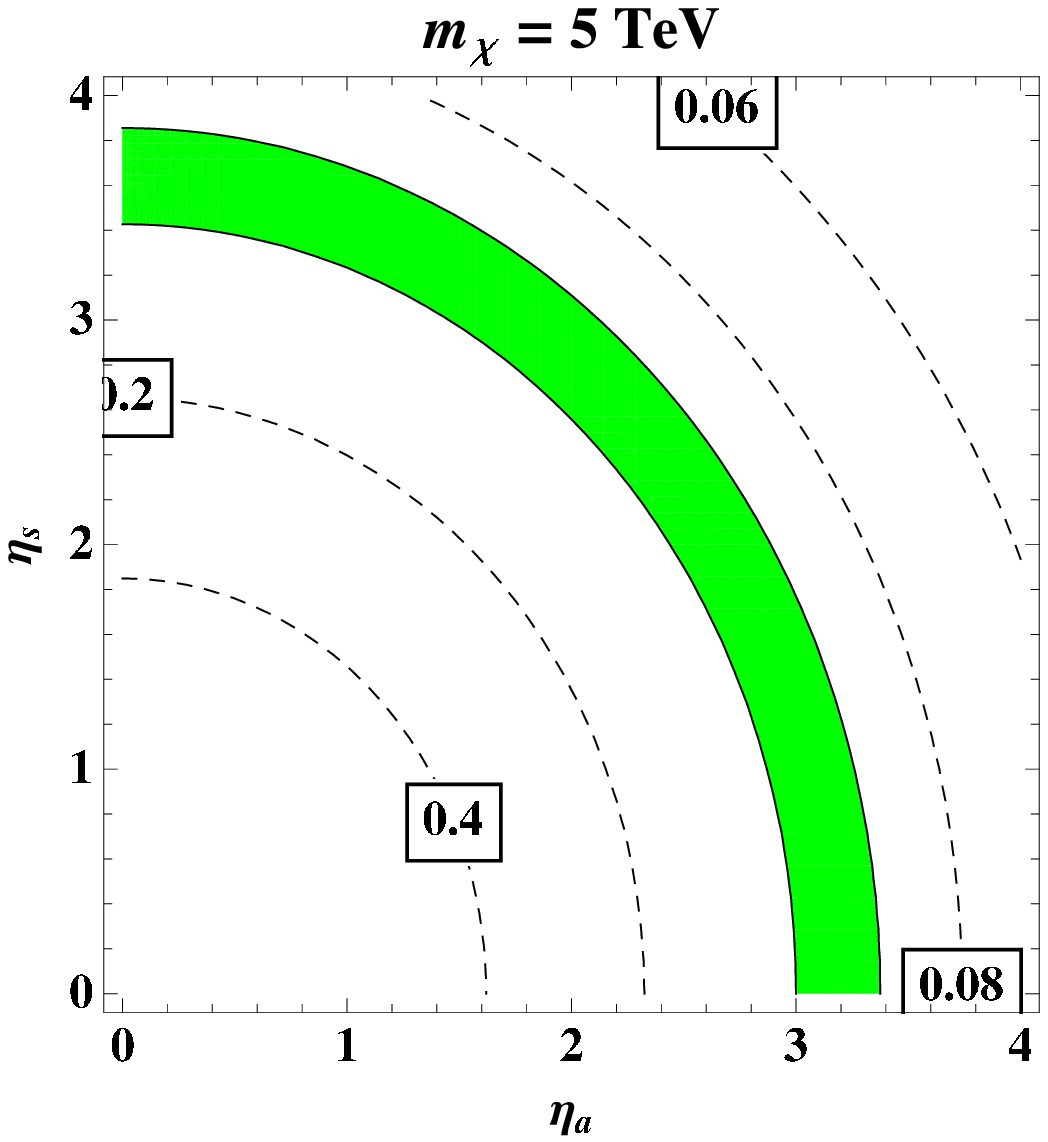} \\ [0.05cm]
\mbox{\bf (a)} & \mbox{\bf (b)}
\end{array}$
\caption{Relic density for the $\chi$ model with $Z_3$ symmetry in the $\left(\eta_a, \eta_s\right)$ plane for relic particle mass: ({\rm a}) $m_{\chi} = 1 \,{\rm TeV}$; ({\rm b}) $m_{\chi} = 5\, {\rm TeV}$. The shaded regions correspond to $0.0975 \leq \Omega_{\chi} h^2 \leq 0.1223$ (WMAP $95\%$ CL region \cite{Dunkley:2008ie}). The dashed lines corresponds to constant dark matter density, and they are along the curves $4 \eta_a^2 + 3 \eta_s^2 \simeq 1.5$.}
\label{fig:chichichiphi}
}

\section{Semi-annihilation with multiple species}
\label{sec:semimulti}
The previous toy model shows that semi-annihilation does affect the dark matter relic density.   However, single-component dark matter with a $Z_3$ symmetry is not representative of the kinds of models in which semi-annihilation is relevant.  In this section, we introduce an example multi-component model having semi-annihilations among its allowed reactions.  A more realistic construction is given in \App{app:SUSYQCD}, where a supersymmetric gauge theory with $N_f = N_c + 1$ is considered.  Here, we prefer to focus on a simpler model with a similar symmetry structure, thus avoiding unnecessary complications caused by having superpartners. 

\subsection{A meson-baryon system}
\label{sec:mesonbaryon}
The dark matter model we consider consists of fermions and bosons that have a ``baryon'' number symmetry as well as a large ``flavor'' symmetry.  The system contains $N_f$ dark baryons, namely vector-like fermions $b^i$ and $b^c_i$, where $i$ is a flavor index.\footnote{We work in two-component notation, where $b$ and $b^c$ are independent left-handed Weyl spinors.}  In addition, there are $N_f^2$ dark mesons, namely complex scalars $\chi_i^j$ arranged in an adjoint of $U(N_f)$.  The charged dark baryons and  mesons interact with a neutral portal field $\phi\equiv {\rm Tr}\left[\chi^j_{i}\right]$, as in \Sec{sec:Z3}.  The particle content and symmetry structure is summarized in \Tab{tab:bbchi}. The Lagrangian of this system is
\begin{align}
\mathcal{L}_{b\bar{b} \chi} &= i \overline{b_i} \bar{\sigma}^{\mu}\partial_\mu b^i + i \overline{b^{c\,i}} \bar{\sigma}^{\mu}\partial_\mu b^c_i + \left|\partial_\mu \chi^j_{i}\right|^2 \nn \\ 
& \quad - \left[m_{b_{i}} \, b^c_i b^i + \lambda\, b^c_j \chi^j_{i} b^i + {\rm h.c.}\right] - V(\chi^j_{i}),
\label{eq:BBbarMLag}
\end{align}
where the scalar potential $V(\chi^j_{i})$ contains all the renormalizable operators involving the meson matrix field consistent with the flavor symmetry in \Tab{tab:bbchi}. In particular it contains the interactions for the singlet $\phi \equiv {\rm Tr}\left[\chi^j_{i}\right]$ analogous to the ones in \Eq{eq:Vphi}, allowing such a singlet to decay to SM particles.
\label{sec:BBbarM}

\TABLE[t]{
\parbox{5in}{ 
\begin{center}
\begin{tabular}{|c|c|cc|}
\hline
 & spin & $U(N_f)$ & $U(1)_B$ \\ 
\hline\hline
$b^i$ & Weyl left & $\overline{\Box}$ & $1$ \\
$b^c_i$ & Weyl left &  $\Box$ & $-1$ \\
$\chi_i^j$ & complex scalar & ${\bf Adj} $ & $0$ \\
\hline
\end{tabular}
\end{center}
}
\caption{Field content and symmetries of the meson-baryon system.}
\label{tab:bbchi}
}

In the limit of an exact $U(N_f)$ flavor symmetry, all the particles except $\phi$ are charged, therefore all of the mesons are degenerate and all of the baryons are degenerate.  In addition, if the triangle inequality rule is satisfied for the decay $\chi_i^j \rightarrow b_i \overline{b^c_j}$, then the charged mesons and baryons are stable, and we can consider all of them as dark matter candidates. If a spurion $\lambda^j_i$ is introduced, with $\langle \lambda^j_i \rangle = \delta_{ij} \lambda_i$, the original $U(N_f)$ flavor symmetry is explicitly broken to $U(1)^{N_f}$. The lightest baryon is still stable by the $U(1)_B$ symmetry. In fact, the heavier baryons might be stable as well. While the flavor symmetry allows $b^i$ to decay to the lighter state $b^c_j$ through the process $b^i \rightarrow \overline{b^c_j} \chi^j_i$, such a decay is forbidden by kinematics if the triangle inequality rule $m_{b_i} < m_{b_j} + m_{\chi^j_i}$ is satisfied (see \Fig{fig:triangle}).  The off-diagonal mesons ($i \neq j$) are still charged, thus the lightest one is absolutely stable. The heavier off-diagonal mesons might be stable as well if their symmetry-allowed decays are kinematically forbidden, namely if they satisfy the triangle inequality. Note that the diagonal mesons do not carry any flavor quantum numbers, thus there is no symmetry which guarantees their stability. We denote them by $\chi^i_i = \phi_i$, and the way they decay to SM fields is through operators analogous to \Eq{eq:Vphi}. 

It is now clear from the symmetries of the theory that the semi-annihilations reactions are indeed present when the stable dark matter particles are thermally produced at the freeze-out, as for example
\begin{equation}
\chi^1_2 \,\chi^3_1\, \rightarrow \,\chi^3_2 \,\phi^1_1, \qquad b^{c\,2} \chi^1_2 \rightarrow b_1 \phi^1_1.
\end{equation}
We now give a complete set of Boltzmann equations which can be solved to get the relic abundance in the case of semi-annihilations, and we also explain why in general a semi-analytical solution is not possible in this case.
\FIGURE[t]{
\includegraphics[width=5.5in]{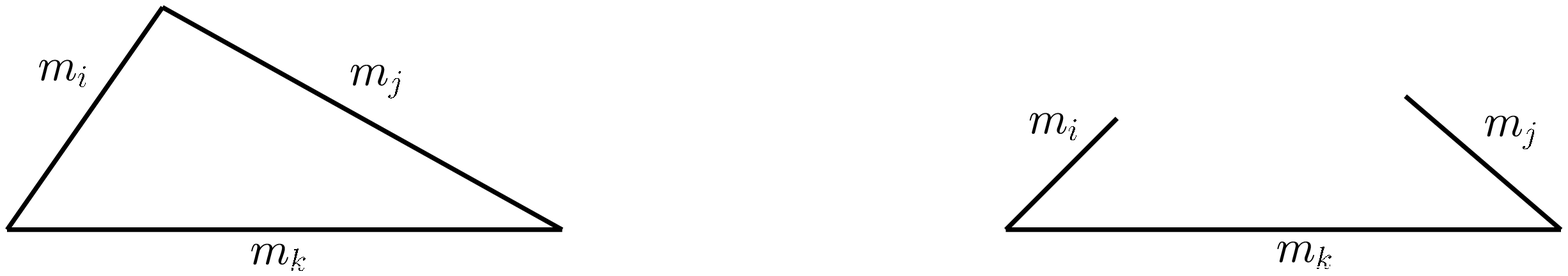}
\caption{Triangle inequality rule for the stability of particles $\psi_i$, $\psi_j$, and $\psi_k$. In the case on the left, the masses can be put on a triangle and thus the three particles are mutually stable, whereas for the case on the right, this is not possible and the decay $\psi_k \rightarrow \psi_i \psi_j\phi$ is kinematically allowed.}
\label{fig:triangle}
}

\subsection{Boltzmann equations}
We consider a generic dark matter model with $N$ stable components and assume that the symmetry structure of the theory allows semi-annihilations to take place at the freeze-out. We ignore unstable states in the dark sector, though in principle they would change the relic density through co-annihilation. Given our assumptions, the reactions which have to be included in the Boltzmann equations are the ones in the last row of the \Tab{tab:reactions}. If we label the stable particles as $\psi_i$, where the index $i$ goes from $1$ to $N$, the Boltzmann equation for the number density $n_i$ of the $\psi_i$ particles result in
\begin{equation}
\begin{split}
\frac{d n_i}{dt} + 3 H n_i = & - \langle \sigma_{ii} v_{{\rm rel}} \rangle \left(n^2_i - n_i^{\text{eq}\,2}\right) 
- \sum_{j,k} \langle \sigma_{ijk} v_{{\rm rel}} \rangle \left(n_i n_j - \frac{n_k}{n_k^{\text{eq}}} n_i^{\text{eq}}n_j^{\text{eq}}\right) \\ & - 
\sum_{j,k,m} \langle \sigma_{ijkm} v_{{\rm rel}} \rangle \left(n_i n_j - n_k n_m \frac{n_i^{\text{eq}} n_j^{\text{eq}}}{n_k^{\text{eq}} n_m^{\text{eq}}} \right),
\end{split}
\label{eq:Boltzsystem}
\end{equation}
where $\sigma_{ii}$ is an ordinary annihilation cross section $\psi_i \psi_i \rightarrow \phi \phi^{\prime}$, $\sigma_{ijk}$ is a semi-annihilation cross section $\psi_i \psi_j \rightarrow \psi_k \phi$, and $\sigma_{ijkm}$ is a dark matter conversion process $\psi_i \psi_j \rightarrow \psi_k \psi_m$.  The collision operators on the right-hand side are derived in \App{app:Boltzmann}. 

The pure annihilation contribution $\sigma_{ii}$ must be diagonal in dark matter flavor space, since if it were not, the heavier particle between $\psi_i$ and $\psi_j$ would be unstable by crossing symmetry.  As long as the triangle inequality rule from \Fig{fig:triangle} is satisfied, then the semi-annihilation process $\sigma_{ijk}$ has no such restriction.  The reason why we need to include dark matter conversion $\sigma_{ijkm}$ is because we care about the total energy density $\rho=\sum_i n_i m_i$, and since all of the particles are stable, such reactions can change $\rho$ if the particle masses are different.  Finally, no reaction of the type $\psi_i \phi \rightarrow \psi_j \phi^{\prime}$ takes place, except for the diagonal one $i=j$, by the same stability argument, and we do not have to take into account the diagonal process since it does not change the total number of $\psi_i$ particles.  The lack of the off-diagonal contributions for this process is the reason why in general we cannot have a semi-analytical solution to the system in \Eq{eq:Boltzsystem}, as we now show.

\subsection{Comparision with co-annihilation}
\label{sec:coann}
The system in \Eq{eq:Boltzsystem} contains $N$ coupled equations of the Riccati type, thus it cannot be solved analytically. In the case $N=1$ it is possible to solve the single equation in two different regimes, match the solutions at freeze-out and get a semi-analytical result for the relic density today, as done in \Sec{sec:Z3}. 

However it is well known that in the co-annihilation case, described in the second row in \Tab{tab:reactions}, we can sum the equations in \Eq{eq:Boltzsystem} to get a single one of the same form as the Lee-Weinberg scenario, even if at the beginning we were dealing with a system of $N$ coupled Boltzmann equations \cite{Griest:1990kh}. The crucial difference between co-annihilation and semi-annihilation scenarios is the presence of off-diagonal $\psi_i \phi \rightarrow \psi_j \phi^{\prime}$ reactions in the former.  Such reactions are much more effective at freeze-out than $\psi_i \psi_j$ (co-)annihilation, since to have co-annihilation, we need two dark matter particles in the initial state and their number density is Boltzmann suppressed at that time, whereas the $\phi$ particles have a relativistic number density. 

A more quantitative way to see this fact is to write down the collision operator for the $\psi_i \phi \rightarrow \psi_j \phi^{\prime}$ reaction
\begin{equation}
\mathcal{C}_{\psi_i \phi \rightarrow \psi_j \phi^{\prime}} = n_i n_{\phi} \langle \sigma v\rangle_{\psi_i \phi \rightarrow \psi_j \phi^{\prime}} 
\left(1 - \frac{n_j}{n_i}\frac{n_i^{\text{eq}}}{n_j^{\text{eq}}}\right).
\end{equation}
Since there are a lot of $\phi$ particles around, this reaction is very effective and it guarantees the condition
\begin{equation}
\frac{n_i}{n_j} = \frac{n^{\text{eq}}_i}{n^{\text{eq}}_j} \, , \qquad \qquad r_i \equiv \frac{n_i}{n} = \frac{n^{\text{eq}}_i}{n^{\text{eq}}}.
\label{eq:GScond}
\end{equation}
Thus in the case of co-annihilation (but not in semi-annihilation) we can assume the condition in \Eq{eq:GScond} before, during, and after freeze-out. This relation between equilibrium and non-equilibrium number densities allows for a considerable simplification of the Boltzmann system. Summing these equations together, we get a single equation governing the total number density $n = \sum_i n_i$ involving an effective annihilation cross section $\sigma_{\text{eff}}$ \cite{Griest:1990kh}
\begin{equation}
\frac{d n}{dt} + 3 H n = - \langle\sigma_{\text{eff}} v_{{\rm rel}}\rangle  \left(n^2 - n^{\text{eq}\,2}\right), \qquad \qquad \sigma_{\text{eff}} \equiv \sum_{i,j} \sigma_{ij} \, r_i r_j,
\label{eq:Boltzeff}
\end{equation}
where $\sigma_{ij}$ is the cross section for the $\psi_i \psi_j$ co-annihilation. Since in standard co-annihilation, all heavier states in the dark matter sector decay to the lightest, $\rho_{{\rm DM}} = m_{{\rm lightest}} n$, and \Eq{eq:Boltzeff} determines the dark matter relic density directly. This equation is identical to the Lee-Weinberg case and thus can be solved by an identical technique. 

When semi-annihilations are present, there is simply no way to guarantee the condition in \Eq{eq:GScond}. In fact such a condition is usually not satisfied as we will see in the next section, and therefore we have to solve the system numerically.

\section{The $b b \chi$ model}
\label{sec:bbchi}

While the meson-baryon system in \Sec{sec:mesonbaryon} is representative of dark matter models exhibiting semi-annihilation, numerically solving the $\mathcal{O}(N_f^2)$ coupled Boltzmann equations is not particularly enlightening.  Here, we study in detail a minimal multi-component dark matter model having semi-annihilations among its allowed reactions.  This can be obtained from the system in \Sec{sec:mesonbaryon} by taking $N_f = 2$ and then ``orbifold'' identifying the flavor and baryon symmetries.

In this toy system, dark matter is composed of two stable components, a complex scalar field $\chi$ with mass $m_\chi$ and a vector-like fermion $b$ and $b^c$ with mass $m_b$:
\begin{equation}
\mathcal{L}_{bb\chi, \,{\rm free}} = \left|\partial_\mu \chi\right|^2 + i \overline{b} \bar{\sigma}^{\mu}\partial_\mu b + i \overline{b^{c}} \bar{\sigma}^{\mu}\partial_\mu b^c
- m^2_\chi \chi^\dag \chi - \left[m_b b^c b + {\rm h.c.}\right].
\end{equation}
We impose a $U(1)$ global symmetry under which $b$, $b^c$, and $\chi$ have charges $+1$, $-1$, and $-2$, respectively. The $U(1)$ invariant interacting Lagrangian for this model is
\begin{equation}
\mathcal{L}_{b b \chi, \,{\rm int}} = \left[\kappa_1 \phi b b^c + \kappa_2\chi b b + \kappa_2 \chi^{\dag} b^c b^c + {\rm h.c.} \right] + \kappa_3 \chi^{\dag} \chi \phi + \kappa_4 \chi^{\dag} \chi \phi^2 - V(\phi),
\end{equation}
where we also introduce a real scalar $\phi$ field, the same portal field as in \Sec{sec:Z3}, which we assume to be in thermal equilibrium in our calculation and decays to SM states. The potential $V(\phi)$ is of the same form as \Eq{eq:Vphi}. 

Fermion number conservation always guarantees that $b$ is stable, whereas stability for $\chi$ requires the triangle inequality rule $m_\chi \leq 2 m_b$ (isosceles triangle). In terms of symmetries, the stability of $b$ is guaranteed by a $Z_2$ subgroup of the original $U(1)$ symmetry under which $b$ and $b^c$ goes to minus themselves with all other fields untouched. Likewise, given the triangle inequality, the decay of $\chi$ to other particles is forbidden by a $Z_4$ subgroup of the original $U(1)$ symmetry, under which $\chi$ goes to minus itself (and $b$ and $b^c$ pick up a factor of $i$ and $-i$, respectively). The model and its symmetries are summarized in \Tab{tab:bbchisym}.
\TABLE[t]{
\parbox{5in}{
\begin{center}
\begin{tabular}{|c|c|c|c|c|}
\hline
Fields & Spin & $U(1)$ charge & $Z_4$ & $Z_2$  \\ 
\hline\hline
$b$ & Weyl left & $+1$ & $+i$ & $-1$ \\
$b^c$ & Weyl left & $-1$ & $-i$ & $-1$ \\
$\chi$ & complex scalar & $-2$ & $-1$ & $+1$ \\
$\phi$ & real scalar & $0$ & $0$ & $0$ \\ 
\hline
\end{tabular}
\end{center}
}
\caption{Field content and symmetries of a minimal multi-component model with semi-annihilation.}
\label{tab:bbchisym}
}

\subsection{Boltzmann equations}
We can now write down and solve numerically the Boltzmann equations for the relic abundance of $b$ and $\chi$ and see how semi-annihilations affect the dark matter relic density. In order to write down the system of coupled Boltzmann equations, we have to identify the processes which change the number of $b$ and $\chi$ fields. We have to be careful here, since both fields are complex and we have an extra degree of freedom (the antiparticle). We will follow the evolution of the number density of the particle only for both species, and once we have their relic abundance, we have just to multiply the result by two to take into account also the contribution from the antiparticles. We write the Boltzmann equations for the variables $n_b$ and $n_{\chi}$, and make the assumption $n_b = n_{\bar{b}}$ and $n_{\chi} = n_{\bar{\chi}}$, as for a standard WIMP. 

The reactions which change the number of $b$ particles are
\begin{equation}
b \bar{b} \rightarrow \phi\phi , \qquad b \bar{b} \rightarrow \chi\bar{\chi} , \qquad b b \rightarrow \phi\bar{\chi} , \qquad b \phi \rightarrow \bar{b}\bar{\chi} , \qquad b \chi \rightarrow \bar{b} \phi,
\end{equation}
whereas the ones which change the number of $\chi$ particles are
\begin{equation}
\chi \bar{\chi}  \rightarrow \phi\phi , \qquad \chi\bar{\chi} \rightarrow b \bar{b} , \qquad \chi \phi \rightarrow \bar{b}\bar{b} , \qquad \chi b \rightarrow \phi\bar{b}.
\end{equation}
If we assume $CP$ invariance, the collision operators for $ b \phi \rightarrow \bar{b}\bar{\chi}$ and $b \chi \rightarrow \bar{b} \phi$ cancel out in the Boltzmann equation for $b$, since they do not change the total number of $b + \bar{b}$. The Boltzmann equations take the form
\begin{equation}
\begin{split}
\frac{d n_b}{d t} + 3 H n_b = \sum_{i}\mathcal{C}^i_b,  \qquad \qquad \frac{d n_{\chi}}{d t} + 3 H n_{\chi} = \sum_j \mathcal{C}^j_{\chi},
\end{split}
\end{equation}
where the sums run over all the collision terms. The collision operators for the reactions which change the number of $b$ particles are
\begin{equation}
\begin{split}
\mathcal{C}_{b \bar{b} \rightarrow \phi\phi} = & - \langle \sigma v_{{\rm rel}}\rangle_{b \bar{b} \rightarrow \phi\phi} \left[n_b^2 - n^{\text{eq}\,2}_b\right], \\ \mathcal{C}_{b b \rightarrow \phi\bar{\chi}} = & - \langle \sigma v_{{\rm rel}}\rangle_{b b \rightarrow \phi\bar{\chi}} 
\left[n^2_b - \frac{n^{\text{eq}\,2}_b}{n^{\text{eq}}_{\chi}} n_{\chi}\right], \\
\mathcal{C}_{b \bar{b} \rightarrow \chi\bar{\chi}} = & - \langle \sigma v_{{\rm rel}}\rangle_{b \bar{b} \rightarrow \chi\bar{\chi}} 
\left[n^2_b - \frac{n^{\text{eq}\,2}_b}{n^{\text{eq}\,2}_{\chi}}n^2_{\chi}\right],
\end{split}
\end{equation}
whereas the ones for the reactions which change the number of $\chi$ particles are\footnote{The collision term for $\chi \phi \rightarrow \bar{b}\bar{b}$  is written by using the opposite reaction, see \Eq{eq:BEinv} in \App{app:Boltzmann}. The factor of $1/2$ comes from the thermal average over two identical particle in the initial state, which is not compensated by any factor of $2$ since in this process we get rid of one single $\chi$ particle.} 
\begin{equation}
\begin{split}
\mathcal{C}_{\chi \bar{\chi}  \rightarrow \phi\phi} = & - \langle \sigma v_{{\rm rel}}\rangle_{\chi \bar{\chi}  \rightarrow \phi\phi} \left[n_{\chi}^2 - n^{\text{eq}\,2}_{\chi}\right], \\
\mathcal{C}_{\chi b \rightarrow \phi\bar{b}} = & - \langle \sigma v_{{\rm rel}}\rangle_{\chi b \rightarrow \phi\bar{b}} \left[n_b n_{\chi} - n_b n^{\text{eq}}_{\chi}\right], \\
\mathcal{C}_{\chi\bar{\chi} \rightarrow b \bar{b}} = & - \langle \sigma v_{{\rm rel}}\rangle_{\chi\bar{\chi} \rightarrow b \bar{b}} 
\left[n_{\chi}^2 - \frac{n^{\text{eq}\,2}_{\chi}}{n^{\text{eq}\,2}_{b}} n_{b}^2\right], \\
\mathcal{C}_{\chi \phi \rightarrow \bar{b}\bar{b}} = & - \frac{1}{2} \langle \sigma v_{{\rm rel}}\rangle_{\bar{b}\bar{b} \rightarrow \chi \phi} 
\left[\frac{n^{\text{eq}\,2}_{b}}{n^{\text{eq}}_{\chi}}n_{\chi} - n_{b}^2\right].
\end{split}
\end{equation}

Once we fix the dark matter masses and limit our consideration to $s$-wave annihilation, there are four free parameters, which we choose to be the following $s$-wave matrix elements
\begin{equation}
\mathcal{M}_{b \bar{b} \rightarrow \phi \phi} = \alpha, \qquad \mathcal{M}_{\chi \bar{\chi} \rightarrow \phi \phi} = \beta, 
\qquad \mathcal{M}_{b \bar{b} \rightarrow \chi \bar{\chi}} = \kappa,  \qquad \mathcal{M}_{\chi b \rightarrow \phi \bar{b}} = \epsilon.
\label{eq:bbchireacs}
\end{equation}
The diagrams for the reactions in \Eq{eq:bbchireacs} are shown in \Fig{fig:bbchiFeyn}, and the matrix elements for the other reactions can be obtained by crossing symmetry. The first two amplitudes come from the standard annihilation to light particles in the final state, the third from the process where two dark matter particles of one species are entirely converted to two dark matter particles of the other species. This process is phase space suppressed for typical kinetic energies at freeze-out. The last corresponds to the semi-annihilation process, which is the new effect we take into account. In particular we are interested to see how the relic abundance changes with the parameter $\epsilon$.  The explicit expression for the thermal averages relevant for the collision operators are given in \App{app:thaverages}.
\FIGURE[t]{
\includegraphics[scale=0.53]{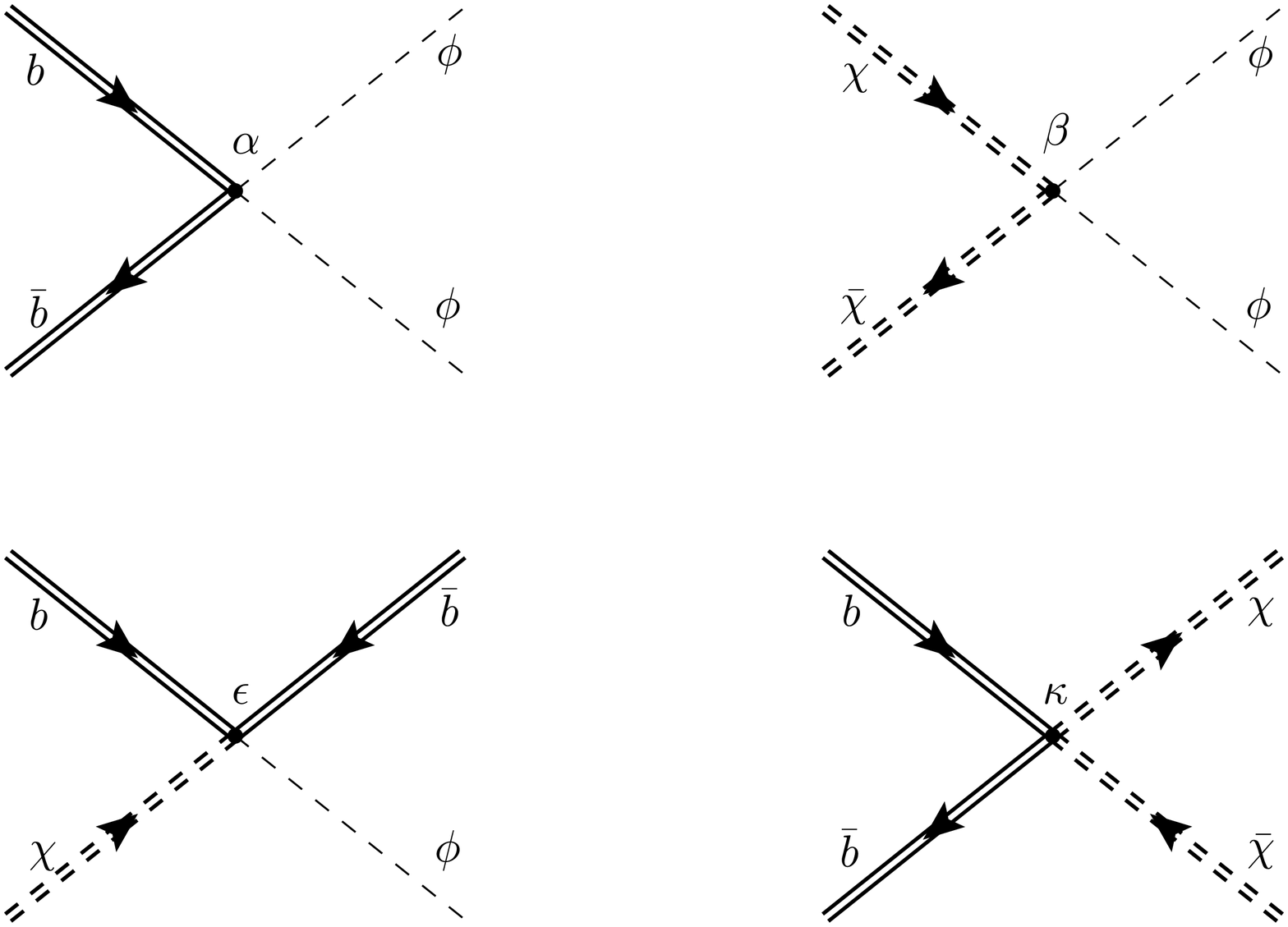}
\caption{Reactions in the $b b \chi$ model. All the other ones can be obtained by crossing symmetry. The dark matter particles are drawn as double lines, and the coefficients of the $s$-wave amplitudes are given in \Eq{eq:bbchireacs}.}
\label{fig:bbchiFeyn}
}

In order to find the relic abundance of $b$ and $\chi$ particles, we have to solve a system of two coupled Boltzmann equations. As discussed in \Sec{sec:coann}, it is not possible to sum the equations and get a single Boltzmann equation, as one can do in the case of co-annihilations. The reason is that the reaction $b \phi \rightarrow \chi \phi$ does not take place at all, since it violates the $U(1)$ symmetry of the model as well as fermion number, thus we cannot make any assumptions on the relative number density of the two species. We have to solve the system numerically, and to do this it is convenient to write this system in dimensionless variables. We define
\begin{equation}
Y_i \equiv \frac{n_i}{s}, \qquad x \equiv \frac{m_\chi}{T}, \qquad \lambda_{ab \rightarrow cd}(x)  \equiv \frac{s(x=1)}{H(x=1)} \langle \sigma v_{{\rm rel}}\rangle_{ab \rightarrow cd}(x) ,
\label{eq:Ylambdadef}
\end{equation}
where the time variable $x$ is defined with respect to the mass of $\chi$.  The Boltzmann equations in the new variables are
\begin{equation}
\begin{split}
& \frac{d Y_b}{d x} = - \frac{1}{x^2} \left\{\lambda_{b \bar{b} \rightarrow \phi\phi} \left[Y_b^2 - Y^{\text{eq}\,2}_b\right] + \lambda_{b \bar{b} \rightarrow \chi\bar{\chi}} 
\left[Y^2_b - \frac{Y^{\text{eq}\,2}_b}{Y^{\text{eq}\,2}_{\chi}}Y^2_{\chi}\right] 
+ \lambda_{b b \rightarrow \phi\bar{\chi}} 
\left[Y^2_b - \frac{Y^{\text{eq}\,2}_b}{Y^{\text{eq}}_{\chi}} Y_{\chi}\right] \right\},
\\
& \frac{d Y_{\chi}}{d x} = - \frac{1}{x^2} \left\{\lambda_{\chi \bar{\chi}  \rightarrow \phi\phi} \left[Y_{\chi}^2 - Y^{\text{eq}\,2}_{\chi}\right] + \lambda_{\chi\bar{\chi} \rightarrow b \bar{b}} 
\left[Y_{\chi}^2 - \frac{Y^{\text{eq}\,2}_{\chi}}{Y^{\text{eq}\,2}_{b}} Y_{b}^2\right] \right. \\ & \left. \qquad \qquad +~\frac{1}{2} \lambda_{\bar{b}\bar{b} \rightarrow \chi \phi} 
\left[\frac{Y^{\text{eq}\,2}_{b}}{Y^{\text{eq}}_{\chi}}Y_{\chi} - Y_{b}^2\right] + \lambda_{\chi b \rightarrow \phi\bar{b}} \left[Y_b Y_{\chi} - Y_b Y^{\text{eq}}_{\chi}\right]\right\}.
\end{split}
\label{eq:bbchiBEsystem}
\end{equation}
where the functions $\lambda_{ab \rightarrow cd}$ also depend on the time variable $x$. The functions $Y^{\text{eq}}_i$ are defined analogously to the non equilibrium ones in \Eq{eq:Ylambdadef}, and they take the form
\begin{equation}
\begin{split}
Y^{\text{eq}}_{\chi} \equiv & \frac{n^{\text{eq}}_{\chi}}{s} = \frac{g_{\chi}}{g_{* s}} \frac{45}{4\pi^4} x^2 K_2[x] \\
Y^{\text{eq}}_b \equiv & \frac{n^{\text{eq}}_b}{s} = \frac{g_b}{g_{* s}} \frac{45}{4\pi^4}\, r^2 x^2 \, K_2[r x]
\end{split}
\end{equation}
where we define $r=m_b/m_{\chi}$ and $K_2[x]$ is the modified Bessel function.

\subsection{Numerical results}

In our numerical study we fix the dark matter masses to be $m_\chi = 0.8 \TeV$ and $m_b = 1 \TeV$ for concreteness, and we vary the matrix elements amplitudes.   We consider three main cases. For each one, we fix the values of $\alpha$ and $\beta$ (annihilation amplitudes) and study how the relic density is affected once we turn on the parameters $\kappa$ ($b \leftrightarrow \chi$ conversion) or $\epsilon$ (semi-annihilation).
\FIGURE[t]{
$\begin{array}{cc}
\includegraphics[scale=0.65]{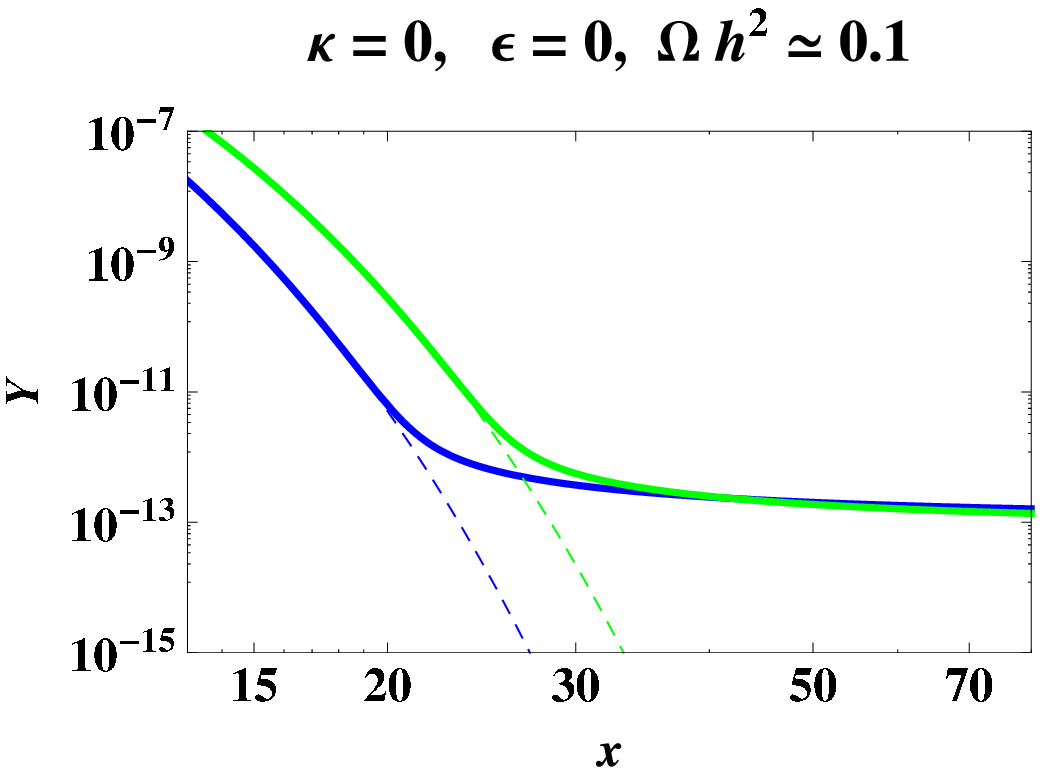} & 
\includegraphics[scale=0.65]{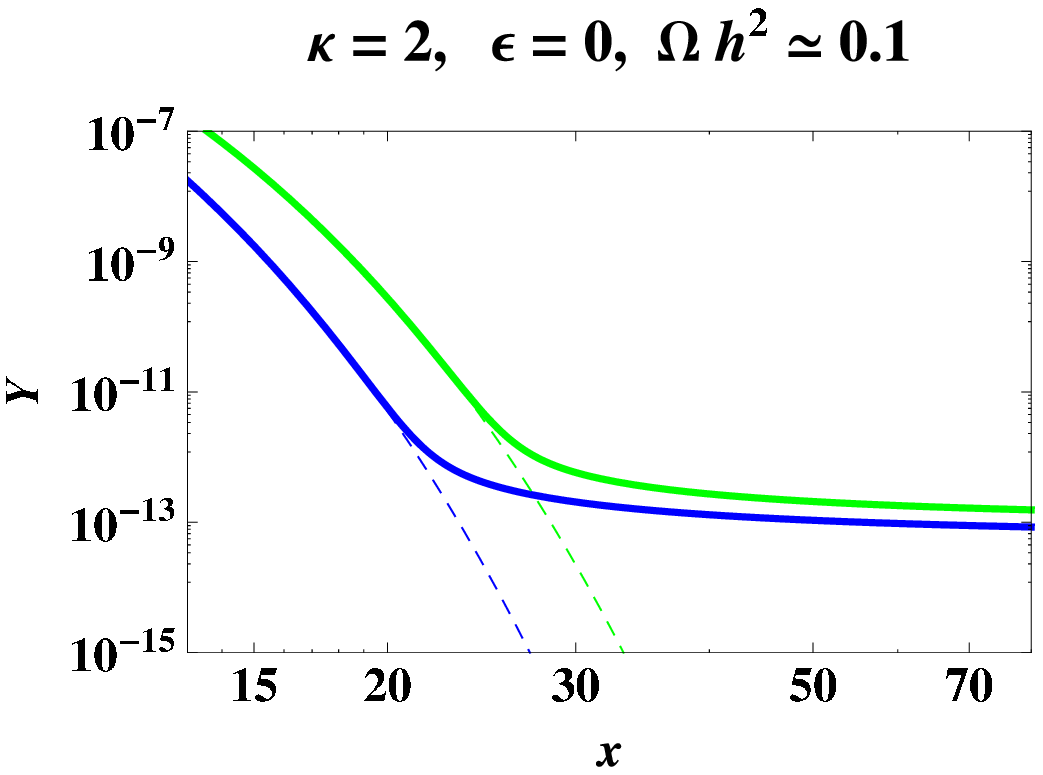} \\
\mbox{\bf (a)} & \mbox{\bf (b)} \\ 
\end{array}$
$\begin{array}{c}
\includegraphics[scale=0.65]{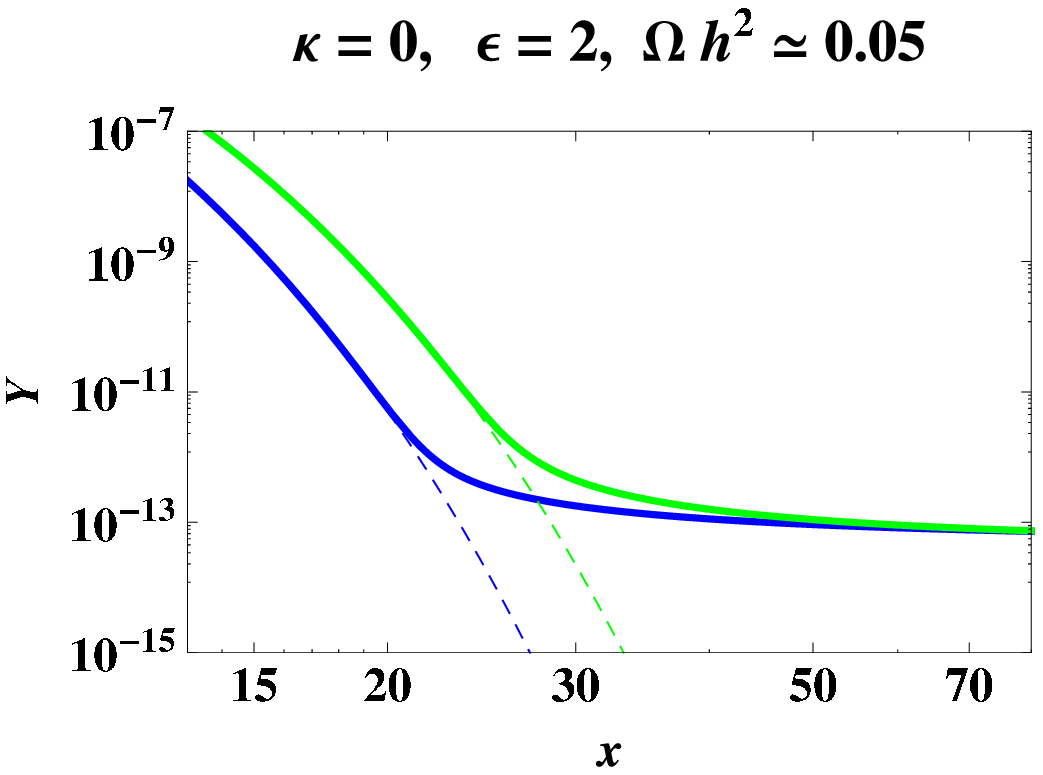}  \\
\mbox{\bf (c)} 
\end{array}$
\caption{$Y$ vs. $x$ evolution in the $bb\chi$ model for the $b$ particles (blue lines) and the $\chi$ particles (green lines). The masses are taken to be $m_\chi = 0.8 \TeV$ and $m_b = 1 \TeV$. We plot the equilibrium values (dashed lines) and the numerical solutions (solid lines). The pure annihilation amplitudes are fixed to be $\alpha = 1.6$ and $\beta = 0.8$, the values of $\kappa$, $\epsilon$ and $\Omega_{{\rm DM}} h^2$ are shown in each plot. We see that both $\kappa$ ($b \leftrightarrow \chi$ conversion) and $\epsilon$ (semi-annihilation) have an effect on the relative and total dark matter densities.}
\label{fig:bbchi1}
}

\FIGURE[t]{
$\begin{array}{cc}
\includegraphics[scale=0.65]{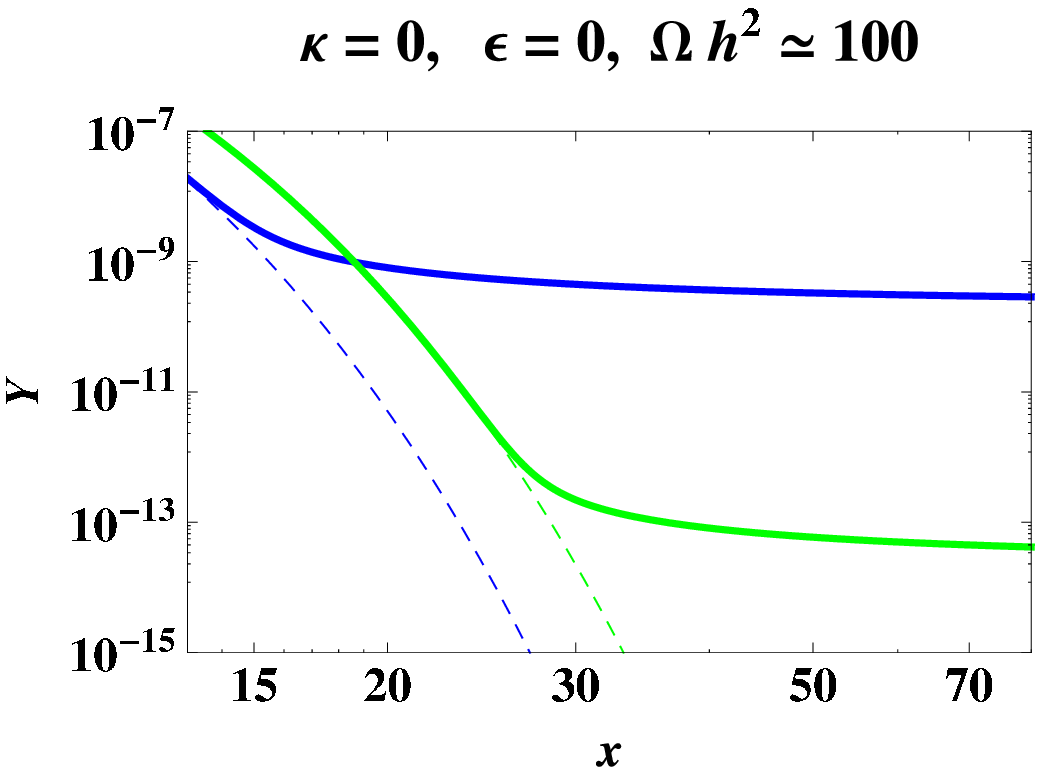} & 
\includegraphics[scale=0.65]{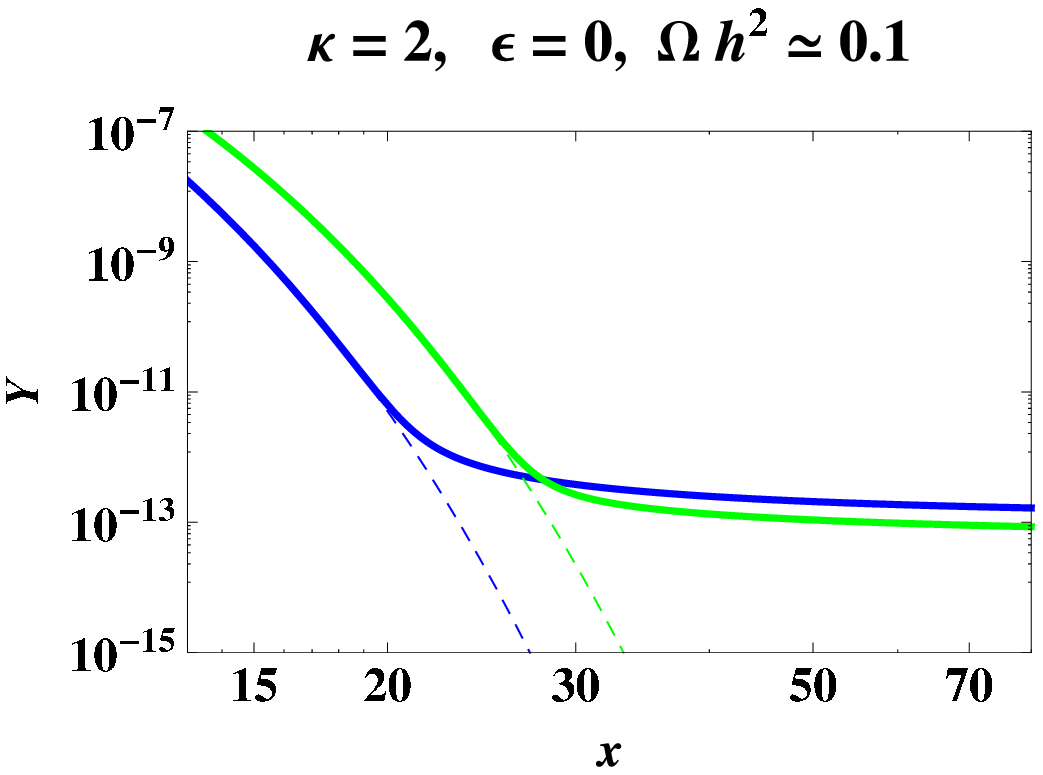} \\
\mbox{\bf (a)} & \mbox{\bf (b)} \\
\end{array}$
$\begin{array}{c}
\includegraphics[scale=0.65]{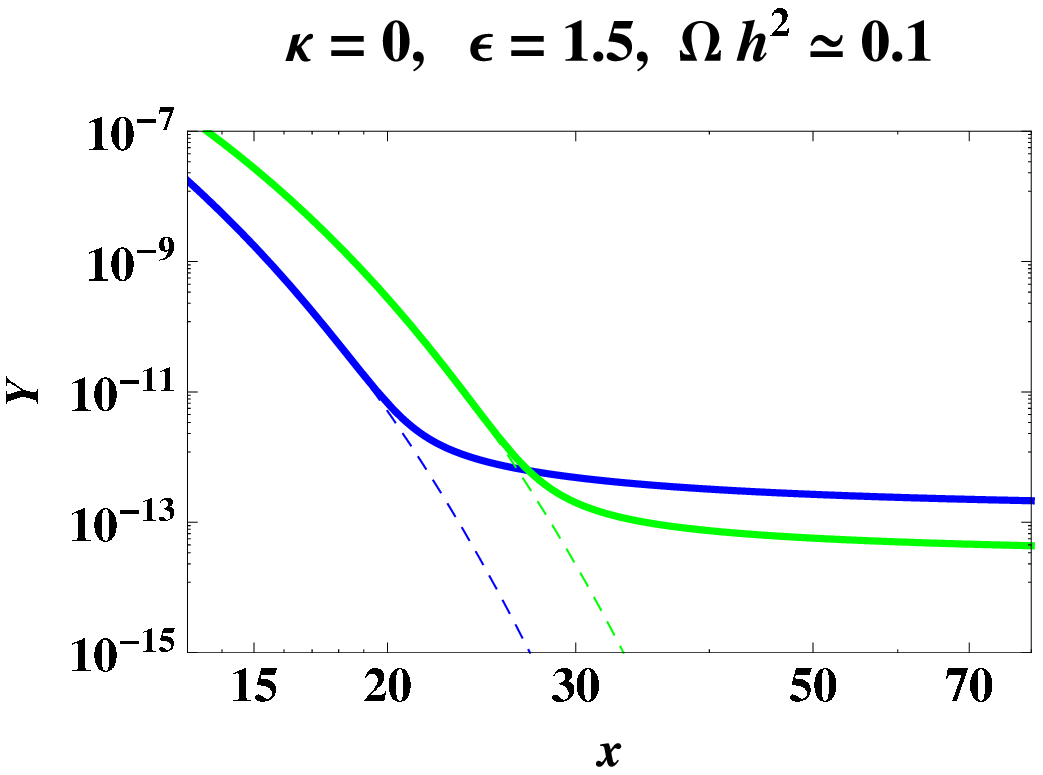}  \\
\mbox{\bf (c)} 
\end{array}$
\caption{Same as \Fig{fig:bbchi1}, now fixing the pure annihilation amplitudes at $\alpha = 0.03$ and $\beta = 1.5$. Without species changing interactions, such a scenario would lead to over production of $b$ particles, but both $\kappa$ ($b \leftrightarrow \chi$ conversion) and $\epsilon$ (semi-annihilation) can be used to achieve the proper relic density.}
\label{fig:bbchi2}
}

\FIGURE[t]{
$\begin{array}{cc}
\includegraphics[scale=0.65]{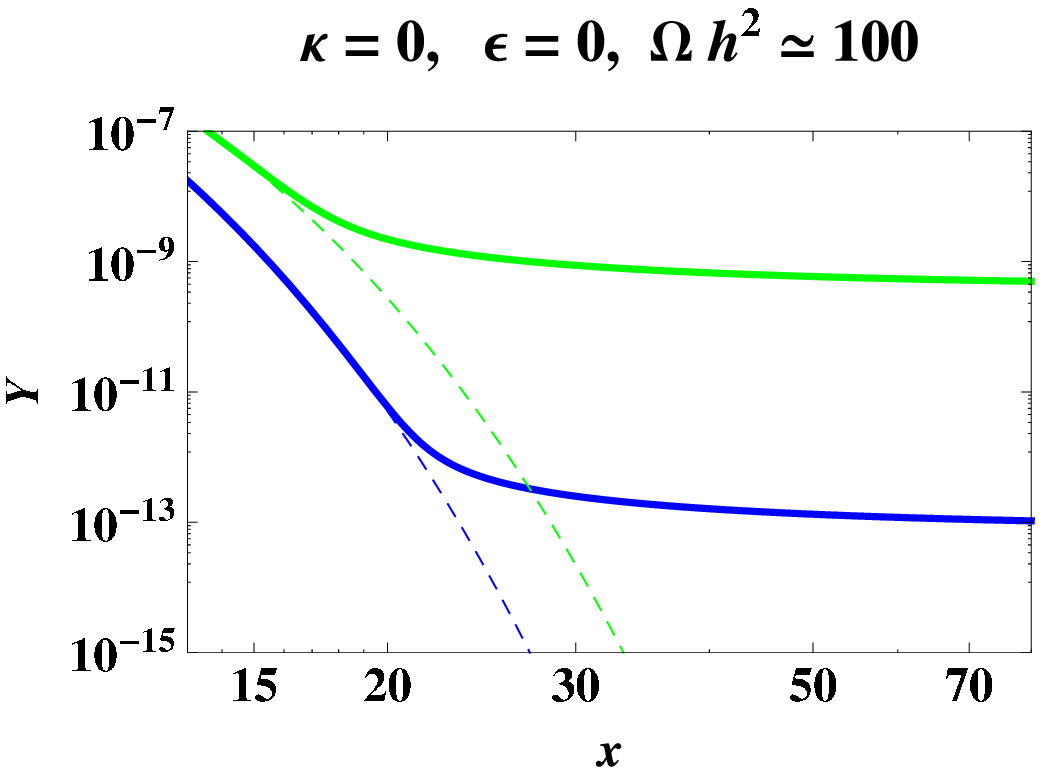} & 
\includegraphics[scale=0.65]{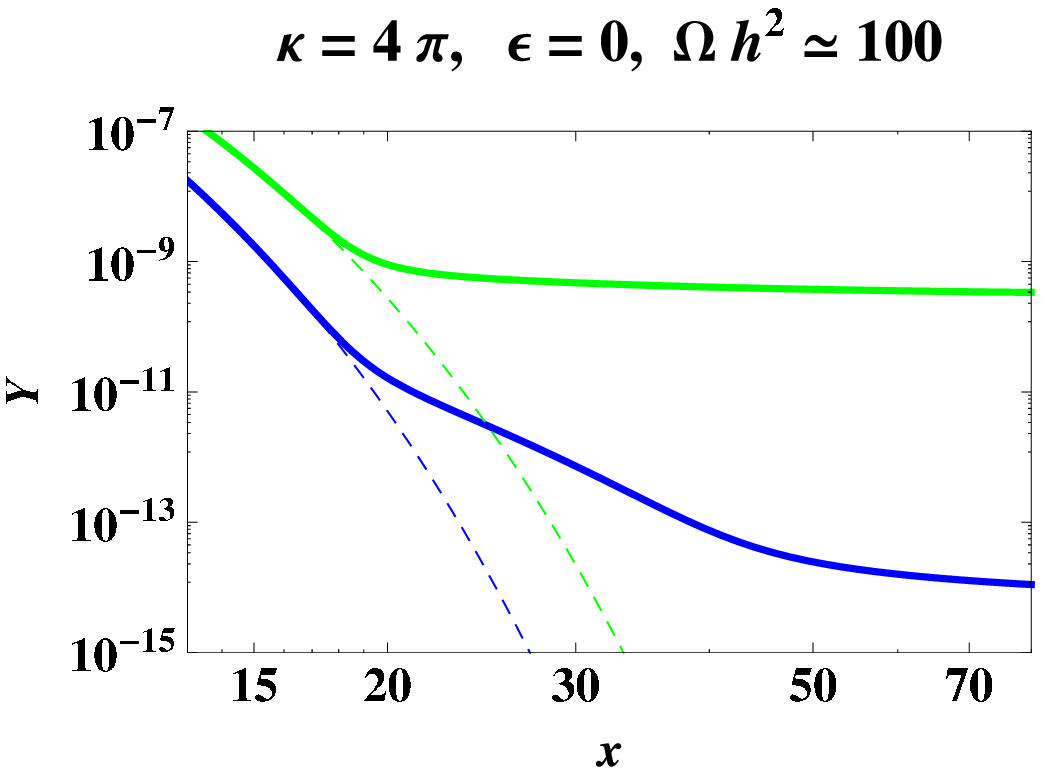} \\
\mbox{\bf (a)} & \mbox{\bf (b)} \\ 
\end{array}$
$\begin{array}{c}
\includegraphics[scale=0.65]{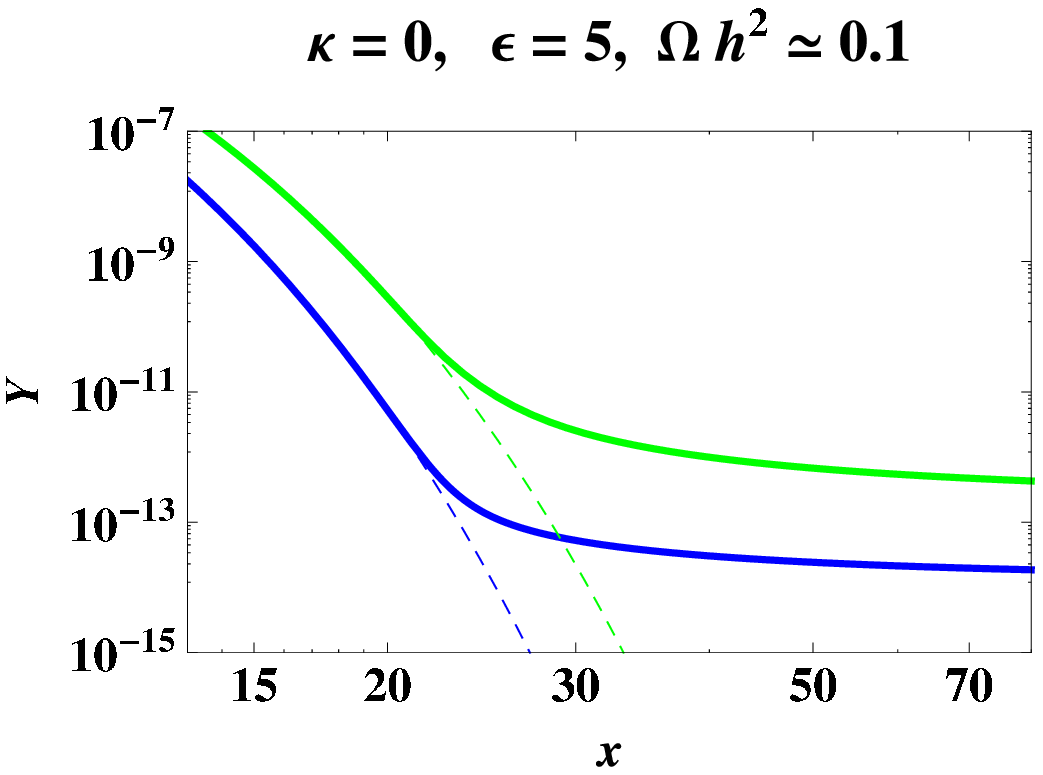}  \\
\mbox{\bf (c)} 
\end{array}$
\caption{Same as \Fig{fig:bbchi1}, now fixing the pure annihilation amplitudes at $\alpha = 2$ and $\beta = 0.01$. Here, $\chi$ particles would be over produced, but because of phase space suppression $\kappa$ ($b \leftrightarrow \chi$ conversion) is ineffective. However, $\epsilon$ (semi-annihilation) is never phase space suppressed and can restore the desired relic density.}
\label{fig:bbchi3}
}

\begin{description}
\item{\bf Case I: $\mathbf{\alpha \simeq \beta}$.}
We start from the case where our choice of $\alpha$ and $\beta$ produces an equal amount of $b$ and $\chi$ today, namely $\alpha = 1.6$ and $\beta = 0.8$, as shown in \Fig{fig:bbchi1}({\rm a}). If we turn on the reaction $b \bar{b} \rightarrow \chi\bar{\chi}$, as in \Fig{fig:bbchi1}({\rm b}), the effect is neglegible and the net result is a small dilution of $b$ particles, as expected, since $b$ is heavier than  $\chi$. If we turn on semi-annihilations, we get fewer relic particles, and for $\epsilon=2$ the relic density is reduced by a factor of $2$ compared to $\epsilon=0$, as shown in \Fig{fig:bbchi1}({\rm c}). Thus we see that semi-annihilations do affect the dark matter relic density.
\item{\bf Case II: $\mathbf{\alpha \ll \beta}$.}
We now consider regions of parameter space where the pure annihilation contributions would create very few particles of one kind and many of the other. In the absence of the amplitudes $\epsilon$ and $\kappa$, such a parameter region would be excluded by overclosure, but the contribution of these species changing reactions can bring the relic abundance back to the observed value. When $\alpha \ll \beta$ the pure annihilations would give many $b$ and very few $\chi$, as shown in \Fig{fig:bbchi2}({\rm a}) for $\alpha=0.03$ and $\beta=1.5$. If we turn $\kappa$ on, as in \Fig{fig:bbchi2}({\rm b}), such a reaction converts $b$ particles into $\chi$ particles very effectively, and the latter have a huge annihilation cross section. The net result is  a reduction of the total dark matter particles $b$ and $\chi$, and for $\kappa=2$ we get the relic abundance we observe today. Analogously, when we turn on the semi-annihilation $b b \rightarrow \phi\bar{\chi}$ we can get the observed relic density, as in \Fig{fig:bbchi2}({\rm c}) for $\epsilon=1.5$, since this reaction destroys $b$ particles to produce $\chi$ particles.
\item{\bf Case III: $\mathbf{\alpha \gg \beta}$.}
From the previous two cases, one might erroneously conclude that $b \leftrightarrow \chi$ conversion and semi-annihilation have similar effects on the relic density. However this is not the case, since $b \leftrightarrow \chi$ conversion is phase space suppressed while semi-annihilation is never phase space suppressed. This is beautifully illustrated in the case when pure annihilations would give a lot of $\chi$ and very few $b$, as shown \Fig{fig:bbchi3}({\rm a}) for $\alpha=2$ and $\beta=0.01$. The semi-annihilation process $\chi b \rightarrow \bar{b} \phi$ is still able to destroy $\chi$ particles and bring the relic density back to the observed value, \Fig{fig:bbchi3}({\rm c}) for $\epsilon = 5$. However, the reaction $\chi\bar{\chi} \rightarrow b \bar{b}$ is now powerless, even if we badly break perturbation theory for $\kappa = 4 \pi$ as in plot \Fig{fig:bbchi3}({\rm b}). This makes sense, since such a reaction is forbidden at zero kinetic energy and at freeze-out, the thermal energy is very small compared with the mass splitting between the two particles. Thus, semi-annihilation is a truly unique species changing interaction that affects early universe cosmology.
\end{description}

We have considered only $m_b > m_{\chi}$ in the above discussion. The results for $m_{\chi} > m_b$ are similar, the main difference being that the conclusions of case II and III are reversed.

\section{Implications for indirect detection}
\label{sec:Indirect}
We have seen how semi-annihilation can affect the predictions for dark matter relic abundance.  Now we explore the implications of semi-annihilations for dark matter detection experiments. Since semi-annihilation does not give any new contributions to direct detection rates, we focus on indirect detection only.\footnote{For an extensive collider study of models with a $Z_3$ symmetry see \Ref{Agashe:2010gt}.  For dual-component WIMP dark matter without semi-annihilation in direct detection, indirect detection, and collider experiments see \Ref{Profumo:2009tb}.}

There has been much recent interest on indirect detection of dark matter, motivated by observations suggesting the presence of a new primary source of galactic electrons and positrons. PAMELA reported an unexpected rise with energy of the $e^+/(e^+ + e^-)$ fraction in comparison with the estimated background \cite{Adriani:2008zr}, supporting what already observed by HEAT \cite{Beatty:2004cy} and AMS-01 \cite{Aguilar:2007yf}. However no antiproton excess was found \cite{Adriani:2008zq}. Such an anomalous excess of the total $e^+ + e^-$ flux was confirmed by FERMI \cite{Abdo:2009zk} and HESS \cite{Collaboration:2008aaa,Aharonian:2009ah}, although the observed spectrum is softer than the peak reported by ATIC \cite{:2008zzr} and PPB-BETS \cite{Torii:2008xu}. These anomalies might be evidence for dark matter annihilation or decay in our galaxy \cite{Cirelli:2008pk,Meade:2009rb,Mardon:2009rc,Meade:2009iu,Ibarra:2009dr}, though astrophysical explanations have been put forward \cite{Hooper:2008kg,Yuksel:2008rf,Serpico:2008te,Profumo:2008ms,Blasi:2009bd}.

In general, semi-annihilation does not lead to the ``boost factor'' necessary to explain the magnitude of these candidate dark matter indirect signals. However the presence of such a reaction can still have considerable implications for indirect searches, as we shall see explicitly for the models considered in \Sec{sec:Z3} and \Sec{sec:bbchi}. Semi-annihilations are additional channels to produce light particles in dark matter interactions in our galaxy, thus the predicted spectrum is enriched with respect to the standard scenarios.  In addition, our examples models can account for the fact that dark matter (semi-)annihilates preferably into leptons; if we assume a portal $\phi$ field mass $m_{\phi} \leq 2 {\rm GeV}$ the decay of $\phi$ to antiprotons is kinematically forbidden \cite{Finkbeiner:2007kk,Pospelov:2007mp,ArkaniHamed:2008qn,Nomura:2008ru}.

\subsection{Cosmic rays via a scalar portal}

In the class of models we consider, semi-annihilations are new reactions to produce light particles in the final state, and for which there is no analog in standard multi-component scenarios.\footnote{There are DM scenarios where DM annihilates into two light states, one unstable and one stable \cite{Chen:2009ab}.  Strictly speaking, this is an example of semi-annihilation, but because the stable final product is much lighter than the DM particles, the kinematics is effectively equivalent to ordinary annihilation.}  We now study the injection spectrum of $\phi$ particles in the models we considered in this paper. As already said, we focus on semi-annihilation to a portal field $\phi$, but in more general dark matter scenarios $\phi$ could be a SM field itself. 

The differential energy flux of $\phi$ from dark matter annihilation and semi-annihilation in the Milky Way is of the form
\begin{equation}
\frac{d \Phi_{\phi}}{d E_{\phi}} \propto \sum_{ij} N^{\phi}_{ij}\, n_i n_j \,\langle \sigma v\rangle_{ij}\, \delta\left(E_{\phi} - E^{ij}_{\phi}\right),
\end{equation}
where the sum runs over the dark matter particles, and $N^{\phi}_{ij}$ and $E^{ij}_{\phi}$ are the number and the energy of $\phi$ produced in any reaction, respectively. Thus, the $\phi$ spectrum consists of a number of monocromatic lines with different intensities. The spectrum of the produced SM particles which we observe in indirect detection experiments would be obtained by convoluting the $\phi$ spectrum with the interactions in \Eq{eq:Vphi}. We consider here as an example the particular case where $\phi$ has a two body decay $\phi \rightarrow \gamma \gamma$. The decay products are isotropic in the $\phi$ rest frame, therefore the energy distribution in the galactic frame is flat. If we assume the mass hierarchy $m_{{\rm DM}} \gg m_{\phi} \gg m_{\gamma}$, the energy distribution of $\gamma$ in the galactic frame has support $0 \leq E_{\gamma} \leq E_{\phi}^{ij}$ for any reaction, and its differential flux takes the form
\begin{equation}
\frac{d \Phi_{\gamma}}{d E_{\gamma}} \propto \sum_{ij} N^{\phi}_{ij}\, \frac{n_i n_j}{E^{ij}_{\phi}} \,\langle \sigma v\rangle_{ij}\; \theta\left(E^{ij}_{\phi}-E_{\gamma}\right).
\label{eq:diffspectrum}
\end{equation}

\subsection{Spectra in $Z_3$ model}
We start with the model with $Z_3$ symmetry considered in \Sec{sec:Z3}, where we have\footnote{The semi-annihilation produces only one $\phi$ particles, however there are two contributions, from $\chi \chi$ and $\bar{\chi}\bar{\chi}$ in the initial state, thus there is no relative factor of $1/2$.} 
\begin{equation}
\Phi_{Z_3} \propto n^2_{\chi} \left[\langle \sigma v \rangle_{\chi\bar{\chi}\rightarrow\phi\phi} + \langle \sigma v \rangle_{\chi\chi\rightarrow\bar{\chi}\phi}\right].
\end{equation}
First, we note that the integrated flux of $\phi$ particles does not differ from the standard case where the thermal production is dominated by annihilations only, namely $\Phi_{Z_3}^0 \propto n_0^2 \langle \sigma v\rangle_0$. The relic density constraint imposes $n_{\chi} = n_{0}$, and it also imposes the equality between the $\langle \sigma v\rangle_0$ and $\langle \sigma v \rangle_{\chi\bar{\chi}\rightarrow\phi\phi} + \langle \sigma v \rangle_{\chi\chi\rightarrow\bar{\chi}\phi}$ (see \Eq{eq:Omegah2Z3}), therefore $\Phi_{Z_3}=\Phi_{Z_3}^0$. Although semi-annihilations cannot provide us with any boost in the indirect detection rate, they do change the injection spectrum of $\phi$ particles. 

In the pure annihilation case, all the $\phi$ particles from dark matter annihilation are produced with energy equal to the mass $m_{\chi}$ of the $\chi$ particle, but in the semi-annihilation reaction, $\phi$ in the final state has energy $3m_{\chi}/4$. In the most general case we have both contributions, thus the $\phi$ injection spectrum has two monochromatic lines at $3m_{\chi}/4$ and $m_{\chi}$, and the relative amplitude of such peaks are given by the ratio of the squared amplitudes. These two quantities are related by the relic density constraint found in \Sec{sec:Z3}, $4\eta_a^2+3\eta_s^2 \simeq 1.5$, thus, to a good approximation, the two ratios depend on only one amplitude, namely
\begin{equation}
R^{\phi}_{m} \equiv \frac{\Phi^{\phi}_{m}}{\Phi^{\phi}_{m}+\Phi^{\phi}_{3m/4}}= 3 \frac{1 - 2 \eta_s^2}{3 + 2 \eta_s^2}, \qquad \qquad 
R^{\phi}_{3m/4} \equiv \frac{\Phi^{\phi}_{3m/4}}{\Phi^{\phi}_{m}+\Phi^{\phi}_{3m/4}}= \frac{8\,\eta_s^2}{3 + 2 \eta_s^2},
\label{eq:Rs}
\end{equation}
where $\Phi^\phi_E$ is the flux of $\phi$ particles with energy $E$.  The results for the $\phi$ spectrum are shown in \Fig{fig:specZ3}.
\FIGURE[t]{
$\begin{array}{cc}
\includegraphics[scale=0.6]{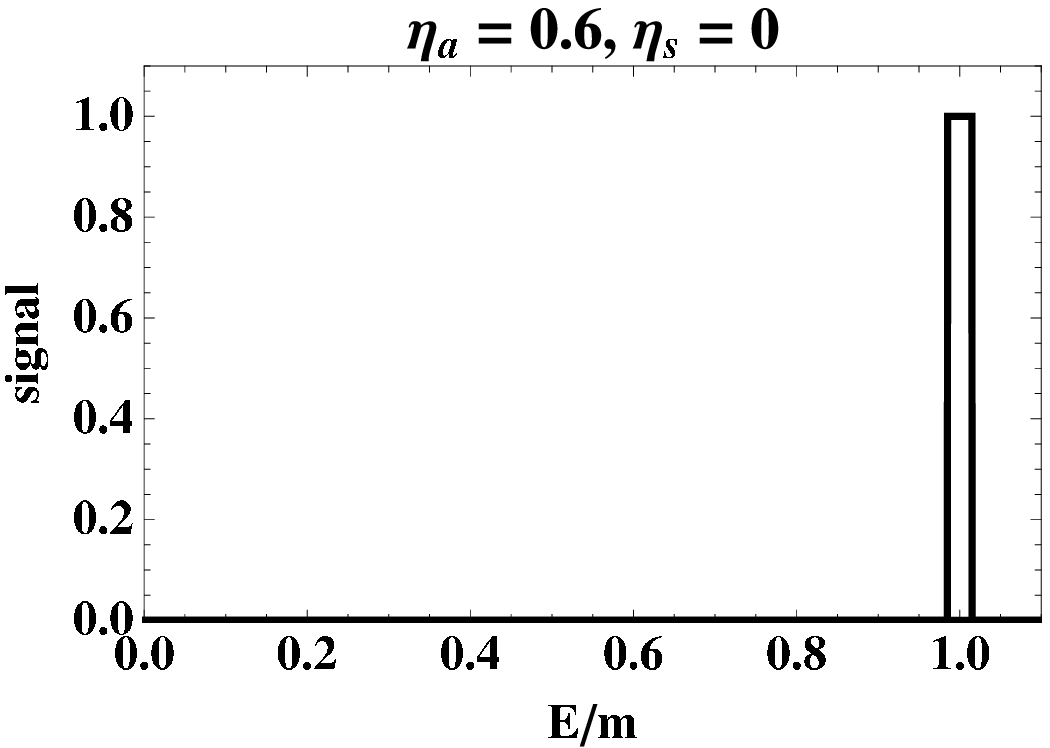} & \hspace{0.4cm} \includegraphics[scale=0.6]{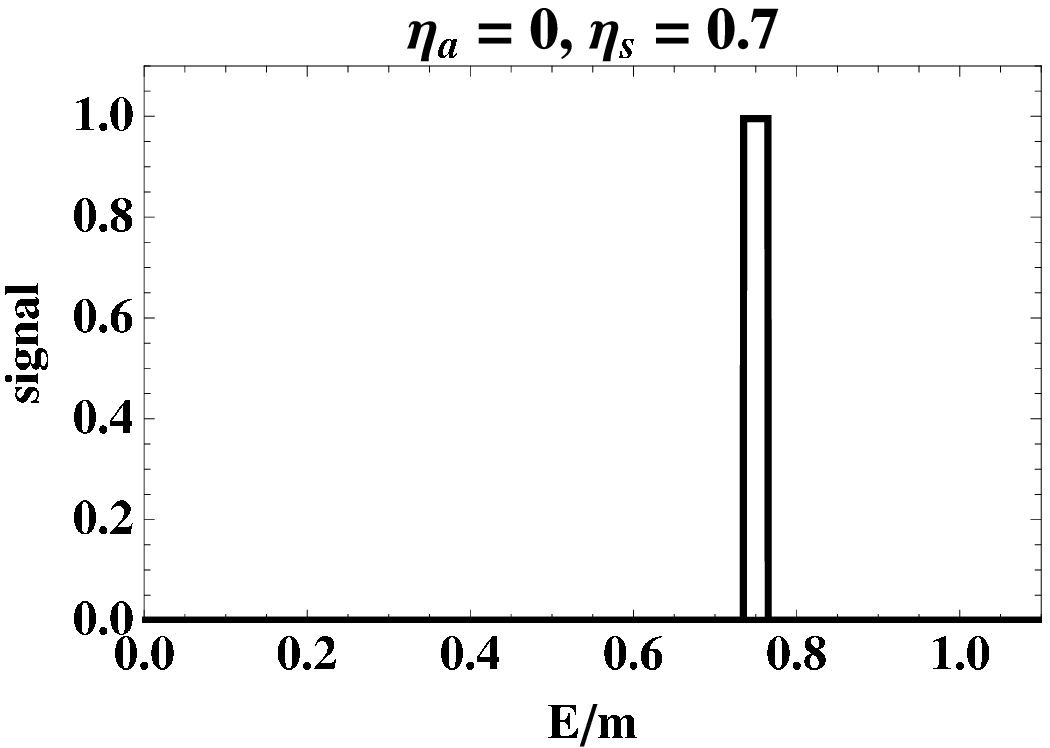} \\ 
\mbox{\bf (a)} & \mbox{\bf (b)} \\[0.3cm]
\includegraphics[scale=0.6]{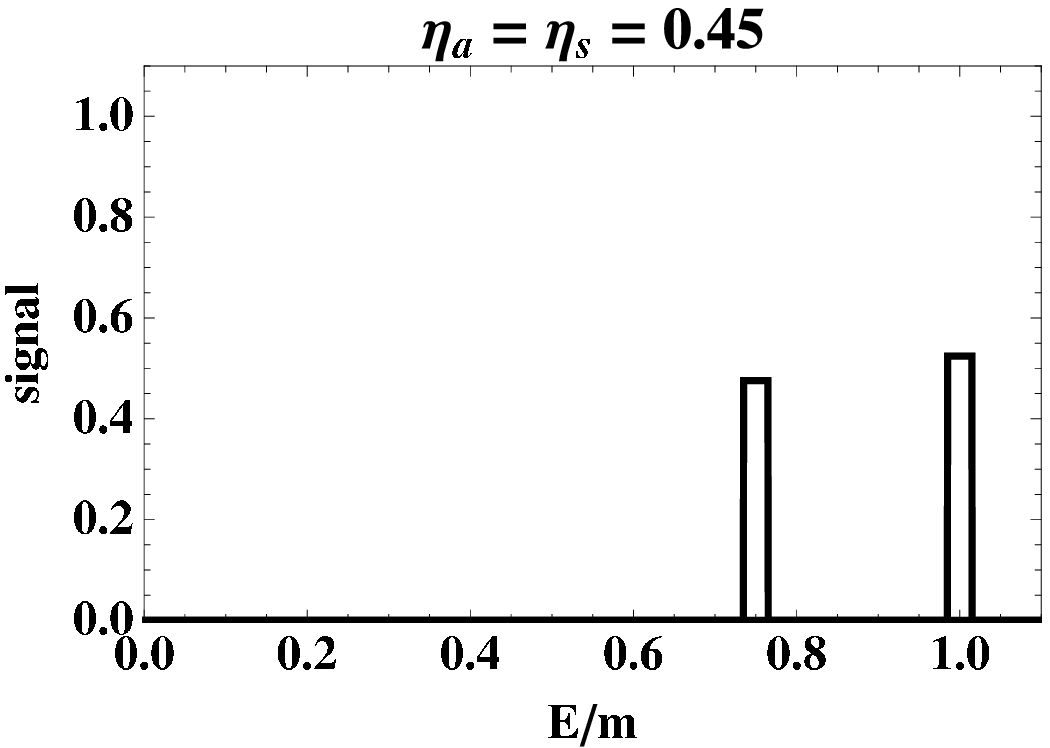} & \hspace{0.4cm} \includegraphics[scale=0.6]{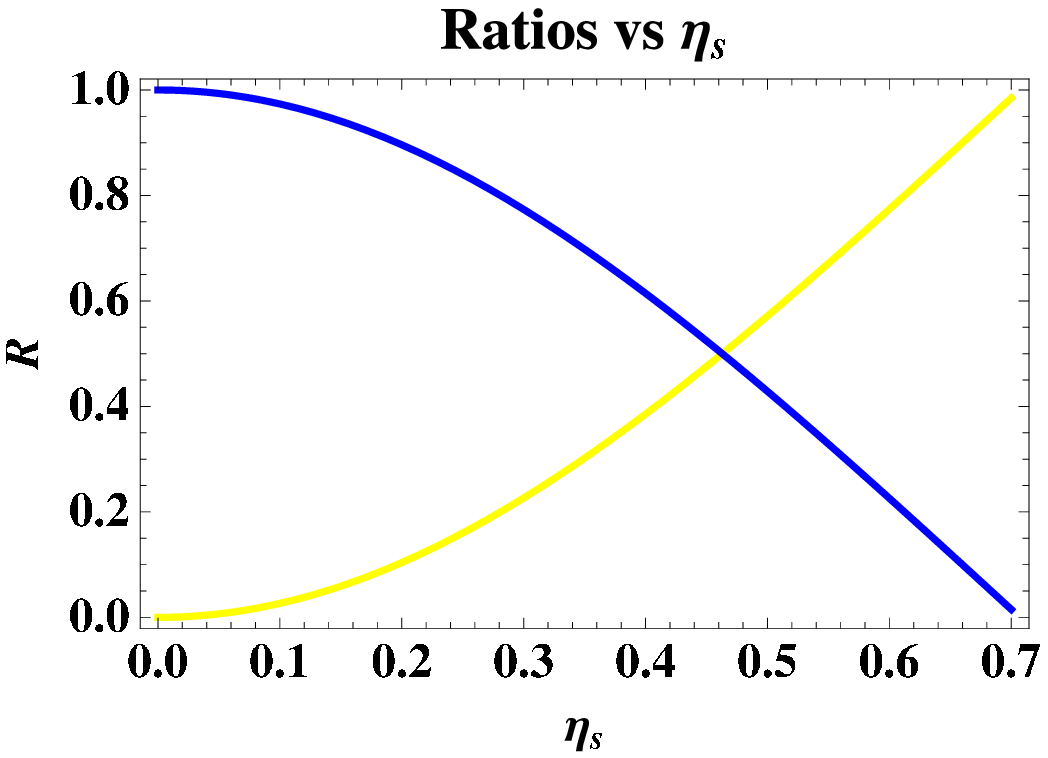} \\
\mbox{\bf (c)} & \mbox{\bf (d)} \\
\end{array}$
\caption{Injection spectrum of $\phi$ particles in the model with $Z_3$ symmetry for: ({\rm a}) annihilation only; ({\rm b}) semi-annihilation only; ({\rm c}) equal amplitudes from the two processes. The total spectrum in normalized to $1$. The relative line intensities as a function of $\eta_s$ are shown in ({\rm d}), $R^{\phi}_{m}$ (blue line) and $R^{\phi}_{3m/4}$ (yellow line). The results are independent of the particle mass $m_{\chi}$.}
\label{fig:specZ3}
}
In the limiting case where only one process is present, the $\phi$ spectrum is monochromatic, see \Fig{fig:specZ3}({\rm a}) and \Fig{fig:specZ3}({\rm b}). When both processes contribute, the spectrum is in general bipolar, as shown in \Fig{fig:specZ3}({\rm c}) for the case of equal amplitudes for the two reactions. The relative spectrum amplitudes in \Eq{eq:Rs} as a function of $\eta_s$ are shown in \Fig{fig:specZ3}({\rm d}). In the case of only one relic particle, the results are independent of the mass $m_{\chi}$ (up to a small logarithmic correction through the freeze-out value in \Eq{eq:xfZ3}).

\subsection{Spectra in $b b \chi$ model}
We now analyze indirect detection in the $bb\chi$ model, which exhibits a much richer kinematic structure.  In this case, there are four reactions to produce $\phi$ particles in the final state, the ordinary annihilations of the two components and two semi-annihilations. The overall $\phi$ flux now takes the form
\begin{equation}
\Phi_{bb\chi} \propto n_b^2 \left[\langle \sigma v_{{\rm rel}}\rangle_{b\bar{b} \rightarrow \phi\phi}+\langle \sigma v_{{\rm rel}}\rangle_{bb \rightarrow \phi\bar{\chi}}\right] + 
n_b n_{\chi} \langle \sigma v_{{\rm rel}}\rangle_{b\chi \rightarrow \bar{b} \phi} + n^2_{\chi} \langle \sigma v_{{\rm rel}}\rangle_{\chi\bar{\chi} \rightarrow \phi \phi}.
\label{eq:bbchiflux}
\end{equation}

Let us first briefly discuss the possibility of achieving a boost factor in this dual-component model. We numerically evaluated the flux in \Eq{eq:bbchiflux} for the region of parameter space which gives the correct relic abundance, and compared such a flux with the one obtained when only the lightest particle is present (this is a conservative comparison, since the flux in the latter case is proportional to $m^{-2}$). We found that the most we can get is a boost factor of a few in the total indirect detection rate. Of course, semi-annihilation could coexist with an alternative boost mechanism such as the Sommerfeld enhancement \cite{Hisano:2004ds,Cirelli:2007xd,ArkaniHamed:2008qn} or a local dark matter overdensity \cite{Hooper:2008kv,Kuhlen:2009is,Yuan:2009bb,Elahi:2009bv}.

The spectrum is this multi-component case is quite rich, since there are now four processes giving $\phi$ in the final state, yelding four monochromatic $\phi$ lines. The energies of the four lines are given by the energies of $\phi$ in the final state
\begin{equation}
E_{b\bar{b} \rightarrow \phi\phi} = m_b,\hspace{0.4cm} E_{bb \rightarrow \phi \bar{\chi}} = m_b - \frac{m_{\chi}^2}{4 m_b}, \hspace{0.4cm}
E_{b\chi \rightarrow \bar{b} \phi} = \frac{m_{\chi}}{2} \frac{2m_b + m_{\chi}}{m_b + m_{\chi}}, \hspace{0.4cm} E_{\chi\bar{\chi} \rightarrow \phi \phi}=m_{\chi}.
\end{equation}

We present numerical results for the $\phi$ spectrum for the same values of the dark matter masses chosen in \Sec{sec:bbchi}, namely $m_\chi = 0.8 \TeV$ and $m_b = 1 \TeV$. This gives $E_{b\bar{b} \rightarrow \phi\phi} = 1 \TeV$, $E_{bb \rightarrow \phi \bar{\chi}} = 0.84 \TeV$, $E_{b\chi \rightarrow \bar{b} \phi} = 0.62 \TeV$ and $E_{\chi\bar{\chi} \rightarrow \phi \phi} = 0.8 \TeV$. To simplify the following discussion we perform our analysis for $\kappa=0$. The inclusion of a finite value of $\kappa$ does not change the expression in \Eq{eq:bbchiflux}, since its associated process does not produce any $\phi$ in the final state, but it can change the thermal relic density as discussed in \Sec{sec:bbchi}. Two example spectra are shown in \Fig{fig:bbchispec}, where two opposite cases are considered. In the first case, \Fig{fig:bbchispec}({\rm a}), the semi-annihilation contribution is taken to be very small, thus the pure annihilation lines $E_{b\bar{b} \rightarrow \phi\phi}$ and $E_{\chi\bar{\chi} \rightarrow \phi \phi}$ dominate. When the thermal production is mostly controlled by semi-annihilations, as in \Fig{fig:bbchispec}({\rm b}), the semi-annihilation lines $E_{bb \rightarrow \phi \bar{\chi}}$ and $E_{b\chi \rightarrow \bar{b} \phi}$ overwhelm the standard contributions.
\FIGURE[t]{
$\begin{array}{cc}
\includegraphics[scale=0.65]{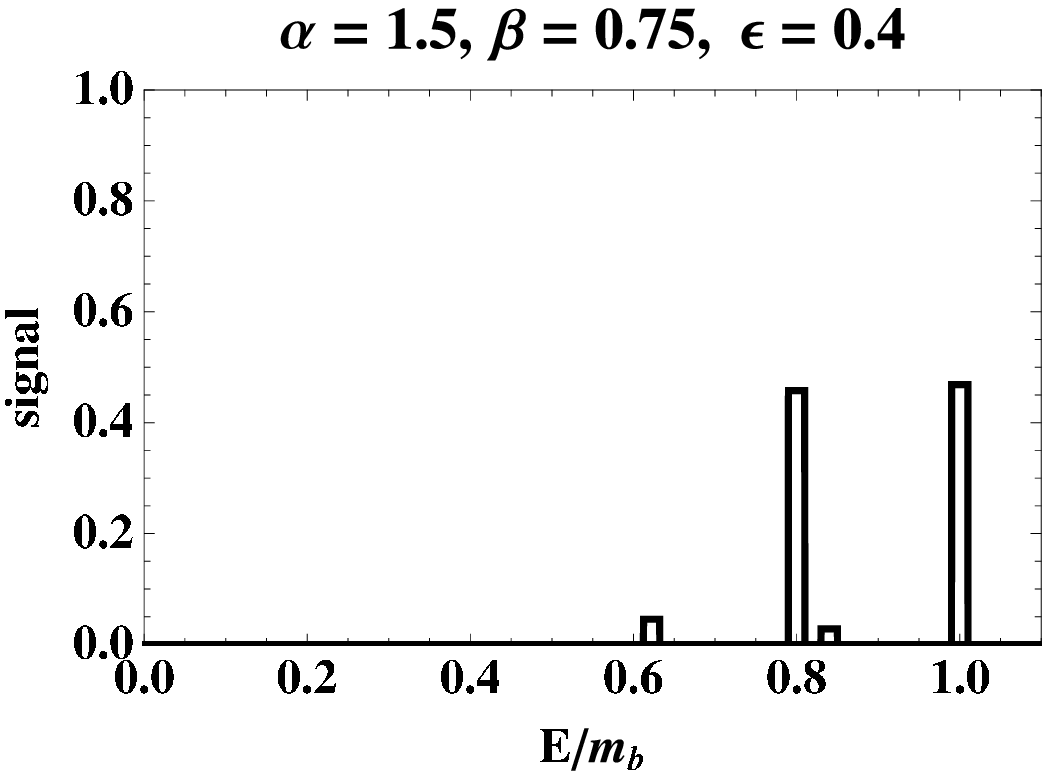} & \hspace{0.5cm}
\includegraphics[scale=0.65]{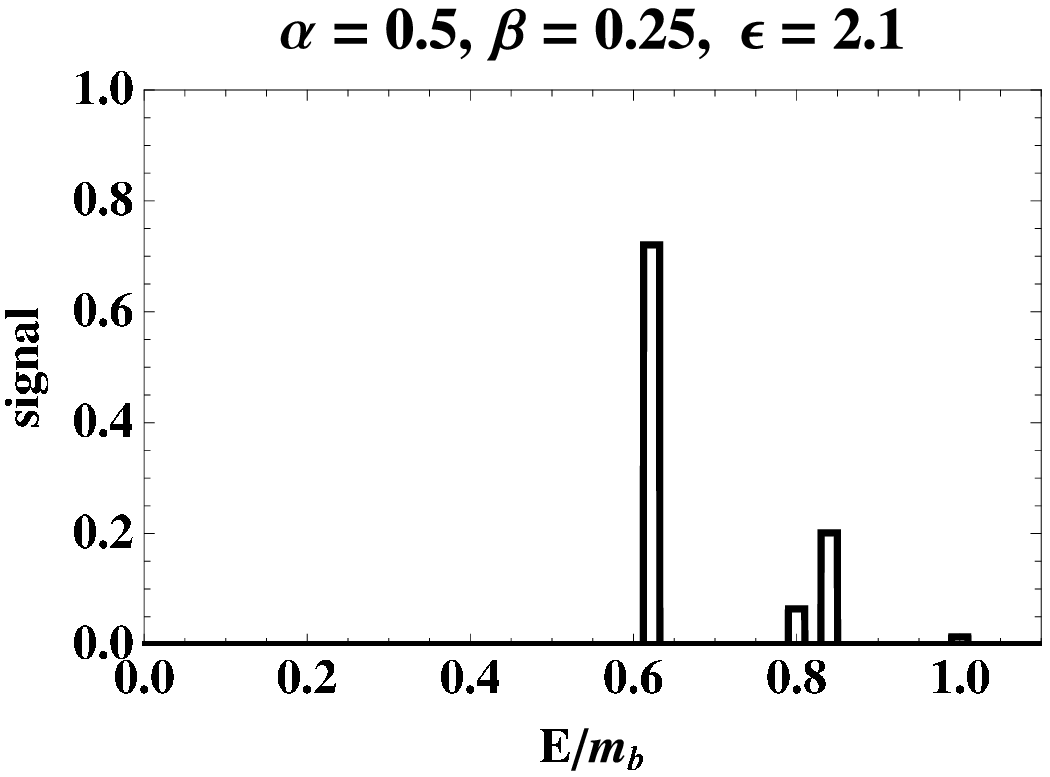} \\
\mbox{\bf (a)} & \mbox{\bf (b)} \\ \vspace{0.3cm} \\
\end{array}$
$\begin{array}{c}
\includegraphics[scale=0.8]{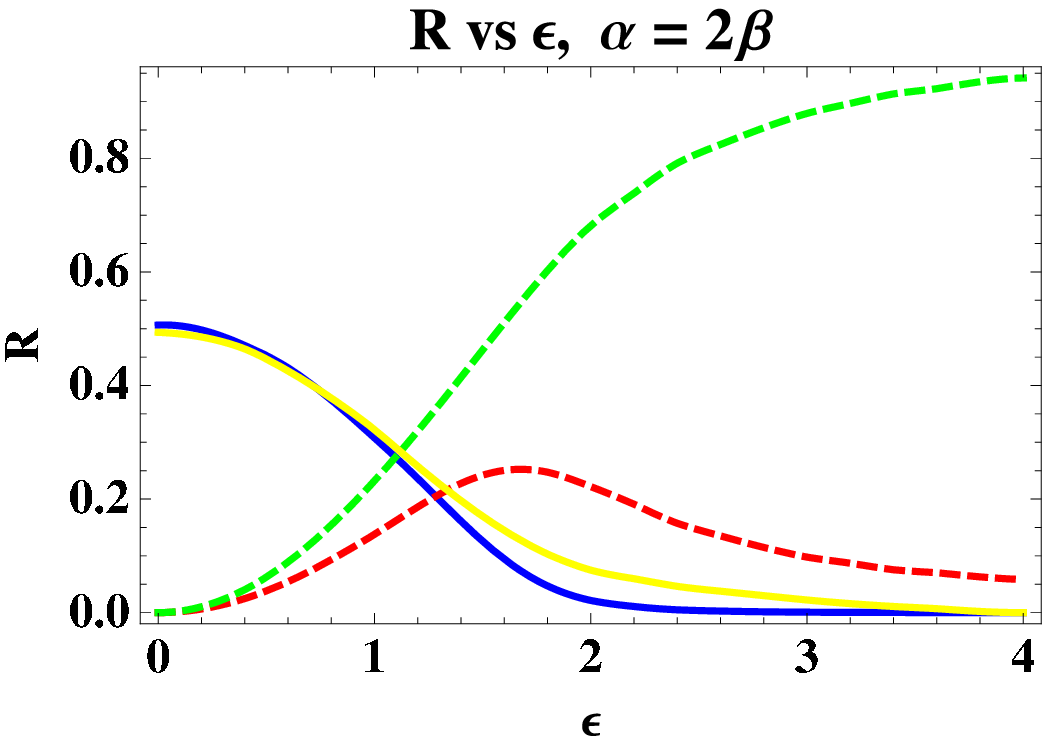}  \\
\mbox{\bf (c)} 
\end{array}$
\caption{Spectra of $\phi$ particles in the $bb\chi$ model for two opposite cases, namely thermal production mostly controlled by: ({\rm a}) annihilations, ({\rm b}) semi-annihilations. The values of the matrix element amplitudes are shown in each plot. The relative intensities of the four lines are shown in ({\rm c}) as a function of $\epsilon$ for $\alpha=2\beta$ keeping the dark matter relic density fixed: $E_{b\bar{b} \rightarrow \phi\phi}$ (blue solid line), $E_{bb \rightarrow \phi \bar{\chi}}$ (red dashed line), $E_{b\chi \rightarrow \bar{b} \phi}$ (green dashed line) and $E_{\chi\bar{\chi} \rightarrow \phi \phi}$ (yellow solid line). We always take $\kappa=0$, $m_\chi = 0.8 \TeV$, and $m_b = 1 \TeV$.}
\label{fig:bbchispec}
}

To study more carefully the relative intensities of the four lines, we fix the ratio $\alpha/\beta = 2$ (as in the two spectra just shown) and keep $\kappa=0$, leaving us with two free parameters, $\alpha$ and $\epsilon$. We take the relic abundance constraint, $\Omega_{{\rm DM}} h^2 \simeq 0.1$, to get $\alpha$ as a function of $\epsilon$, and we study how the line intensities change as we increase the contribution of semi-annihilations. The result is shown in \Fig{fig:bbchispec}({\rm c}). For very small $\epsilon$, the lines from semi-annihilation give a negligible contribution, whereas the standard lines equally share the overall flux (since for $\alpha = 2\beta$ and small $\epsilon$ we get approximately as many $b$ as $\chi$). Once $\epsilon$ gets big enough ($\epsilon \simeq 1$) the thermal production is controlled as much by annihilations as by semi-annihilations, therefore the four lines have approximately equal intensity. For bigger values of $\epsilon$ the annihilation lines intensities are very suppressed, and eventually for $\epsilon \geq 3$ the line with  energy $E_{b\chi \rightarrow \bar{b} \phi}$ dominates the signal. The reason why for large $\epsilon$ only one line is present is easily understood looking the Boltzmann equations system in \Eq{eq:bbchiBEsystem}, and considering only the collision terms associated to semi-annihilations. The collision operator associated to the reaction $\chi \phi \rightarrow \bar{b}\bar{b}$ in the equation for $Y_{\chi}$ tends to increase the total number of $\chi$, thus the dark matter density balance today is totally dominated by the $\chi$ particles, and the line associated to the semi-annhilation with $\chi$ in the initial state gives the dominant contribution. 

One expects the indirect detection spectra in the meson-baryon system in \Sec{sec:mesonbaryon} to have an even richer structure of $\phi$ lines. Of course, once convolved with the $\phi$ decays, the spectrum is less distinct. The injection spectrum when $\phi \rightarrow \gamma \gamma$ (see \Eq{eq:diffspectrum}) is shown in \Fig{fig:bbchidiffspec} for the same two cases considered in \Fig{fig:bbchispec}({\rm a}) and \Fig{fig:bbchispec}({\rm b}). As we increase the semi-annihilation contribution, the differential energy spectrum gets shifted to lower energies. This result is consistent with \Fig{fig:bbchispec}({\rm c}), since $b\chi \rightarrow \bar{b} \phi$ dominates for large $\epsilon$, and such a process has the less energetic line. More detailed study is necessary to know whether such a spectrum can be measured in $\gamma$ rays telescopes.
\FIGURE[t]{
$\begin{array}{cc}
\includegraphics[scale=0.65]{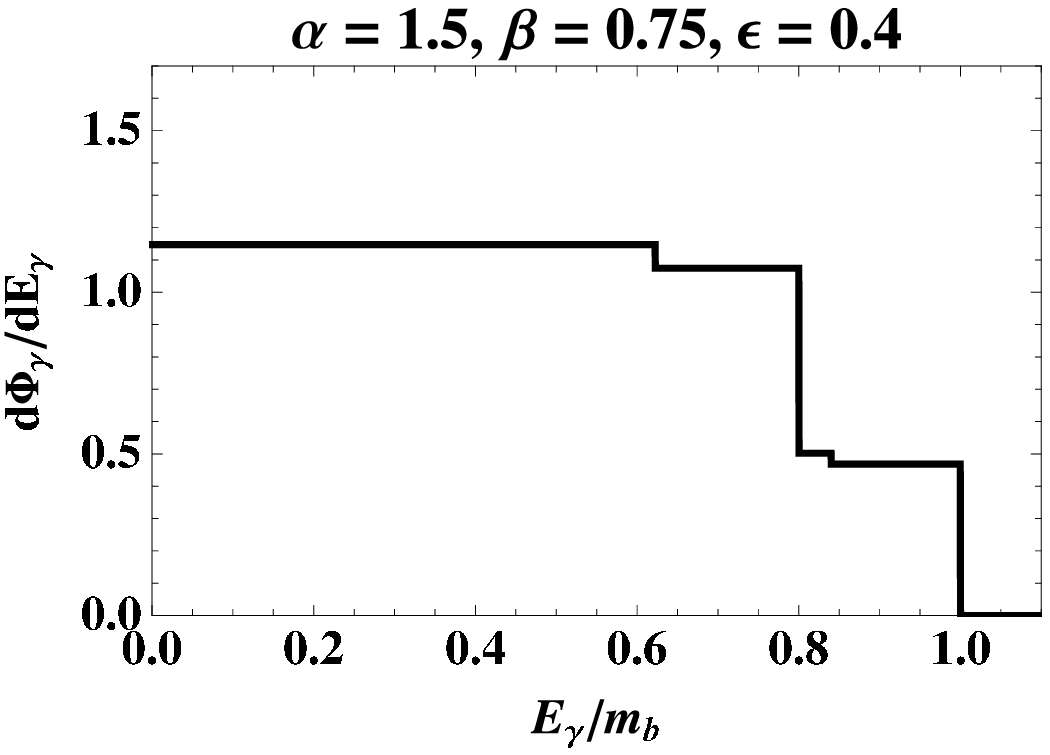} & \hspace{0.5cm}
\includegraphics[scale=0.65]{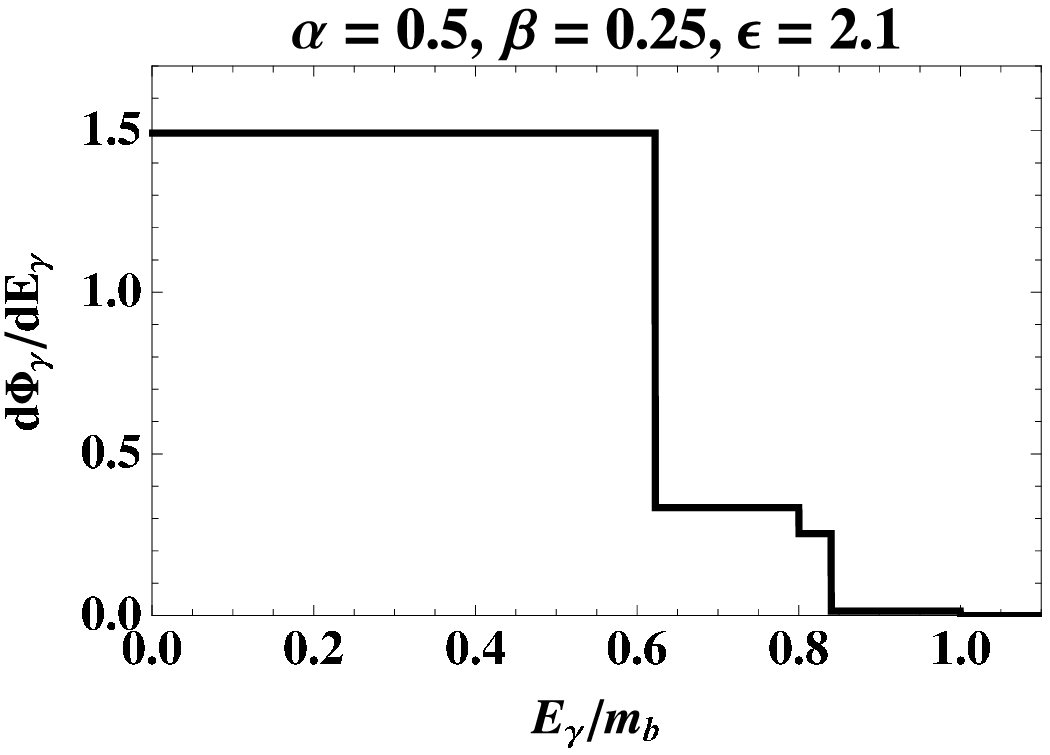} \\
\mbox{\bf (a)} & \mbox{\bf (b)} \\
\end{array}$
\caption{Differential photon energy spectrum in the $bb\chi$ model for the case $\phi \rightarrow \gamma \gamma$. The same two cases as in \Fig{fig:bbchispec} are considered, and the overall signal is normalized to $1$.}
\label{fig:bbchidiffspec}
}

\section{Conclusions}
\label{sec:Conclusions}

The presence of dark matter in our universe motivates us to look for physics beyond the SM, and understanding its origin and composition is one of the great open questions in particle physics and cosmology.  Particle physics models explaining the origin of the Fermi scale often contain stable massive particles called WIMPs.  Assuming WIMPs are thermally produced in the early universe, these particles have the right annihilation cross section to give the observed relic abundance, $\Omega_{{\rm DM}} h^2 \simeq 0.1$, making the WIMP paradigm one of the best solutions to account for dark matter.

In this work, we relaxed some assumptions about the symmetry structure of WIMP interactions by allowing relic particles to ``semi-annihilate''.   We saw that the semi-annihilation process
\be
\psi_i \psi_j \rightarrow \psi_k \phi
\ee
can have a considerable effect on the dark matter relic abundance.  Such processes are present when dark matter has a larger stabilizing symmetry than just $Z_2$, and we showed explicit examples where semi-annihilation is present:  a single species dark matter model with a $Z_3$ symmetry and a multiple species dark matter model with ``baryon'' and ``flavor'' symmetries.  Along with the standard WIMP scenarios summarized in \Tab{tab:reactions}, semi-annihilation does occur in realistic particle physics scenarios.  Indeed, as sketched in \App{app:SUSYQCD}, the simple case of a supersymmetric QCD theory with $N_f = N_c+1$ generically involves semi-annihilation.

We saw that when dark matter is composed of just one species, the effect of semi-annihilation on the relic density can be derived through a standard freeze-out calculation \cite{Lee:1977ua}.  Semi-annihilation can in fact dominate over ordinary annihilation for some regions of parameter space.  However, such single species models are not representation of generic semi-annihilating dark matter models, which usually involve more than one stable dark matter component.  In such cases, a semi-analytical solution for the relic density is simply not possible, and one must resort to numerically solving the complete set of coupled Boltzmann equations given in \Eq{eq:Boltzsystem}.  In particular, the simplifying assumptions made when analyzing co-annihilating models \cite{Griest:1990kh} are not applicable for semi-annihilation.  As a consequence, the dark matter dynamics in the presence of semi-annihilation are far more varied than in decoupled multi-component scenarios.

While the inclusion of semi-annihilation does not give any new contributions to the direct detection of dark matter, it does have interesting implications for indirect detection experiments.  We studied the injection spectrum of light particles in the two toy models we studied, where dark matter (semi-)annihilates through a $\phi$ ``portal'' which subsequently decays to SM states.  The overall integrated flux from (semi-)annihilates in the Milky Way halo is not very affected by semi-annihilation (in the case of one relic particle, is not affected at all).  However, the final state spectrum in semi-annihilating models is far richer than in standards scenarios because of the differing kinematics between semi-annihilation and ordinary annihilation.

In the language of \Ref{Griest:1990kh}, semi-annihilation is in some ways the ``fourth exception'' in the calculation of dark matter thermal relic abundances.  We find it intriguing that unlike the traditional three exceptions (co-annihilation, annihilation below a mass threshold, and annihilation near a pole), this fourth exception not only affects dark matter interactions in the early universe, but also leaves an imprint today via the indirect detection spectrum.  We expect that there are a wide variety of multi-component dark matter models with species changing interactions, motivating further studies of the semi-annihilation process.  

\section*{Acknowledgments}
We thank Kaustubh Agashe for helpful comments.  This work is supported by U.S. Department of Energy (D.O.E.) under cooperative research agreement DE-FG0205ER41360.

\appendix
\section{Boltzmann equations in FRW background}
\label{app:Boltzmann}
In this appendix, we briefly review the derivation of the Boltzmann equations in an expanding universe. If the relic particle, which we assume to be massive, remained in thermal equilibrium until today, its abundance would be negligible because of the exponential factor in the Maxwell-Boltzmann distribution. But if the interactions which keep the particle in thermal equilibrium freeze-out when the temperature is not small compared with the relic particle mass, they can have a significant density today. 

We consider a generic particle $\psi_a$ whose number density evolution is governed by the Boltzmann equation
\begin{equation}
\frac{d n_a}{dt} + 3 H n_a = \mathcal{C}_a,
\end{equation}
where $H$ is the Hubble parameter and $\mathcal{C}_a$ is the collision operator for the reactions which change the number of $\psi_a$ particles. The $3 H n$ factor accounts for the dilution effect of the expansion of the universe, and the right hand side accounts for the interactions that change the number of $\psi$. If there are no interactions, the right hand side is zero and we have $n_a R^3 = {\rm const}$, where $R$ is the scale factor, simply due to the expansion of the universe. 

Considering only $2 \rightarrow 2$ processes, we can write the collision operator as
\begin{equation}
 \mathcal{C}_a = \sum_{b,c,d} \mathcal{C}_{ab \rightarrow cd},
\end{equation}
where the sum runs over all the possible reaction $ab \rightarrow cd$. The collision operator for a single reaction results in
\begin{equation}
\mathcal{C}_{ab \rightarrow cd} = - \int (2\pi)^4 \delta^4\left(p_a+p_b-p_c-p_d\right) \,d\Pi_a d\Pi_b d\Pi_c d\Pi_d \left[\left|\mathcal{M}_{ab \rightarrow cd}\right|^2\,f_a f_b - \left|\mathcal{M}_{cd \rightarrow ab}\right|^2\,f_c f_d\right],
\end{equation}
where $f_i$ is the phase space distribution for the particle $i$. The relativistic invariant phase space is defined as
\begin{equation}
d \Pi_i = g_i \frac{d^3 p_i}{2 E_i (2\pi)^3}.
\end{equation}
where $g_i$ is the number of the internal degrees of freedom of the particle $i$. If we assume $CP$ (or $T$) invariance, we have an equality between the two matrix elements $\mathcal{M}_{ab \rightarrow cd} = \mathcal{M}_{cd \rightarrow ab}$, and we can rewrite
\begin{equation}
\mathcal{C}_{ab \rightarrow cd} = - \int (2\pi)^4 \delta^4\left(p_a+p_b-p_c-p_d\right) \,d\Pi_a d\Pi_b d\Pi_c d\Pi_d\,\left|\mathcal{M}_{ab \rightarrow cd}\right|^2\, \left[f_a f_b - f_c f_d\right].
\end{equation}

In the following, we will also assume kinetic equilibrium during freeze-out
\begin{equation}
f_i(E,t) = \frac{n_i(t)}{n_i^{\text{eq}}(t)}\,f_{i}^{\text{eq}}(E,t),
\end{equation}
with the equilibrium number density being
\begin{equation}
n_i^{\text{eq}} = g_i \int \frac{d^3 p}{(2\pi)^3} f_i^{\text{eq}}(\mathbf{p}),
\end{equation}
where $f_i^{\text{eq}}(\mathbf{p})$ is the thermal distribution for the species when it is in thermal equilibrium. Thus the spectrum remains thermal up to a constant which takes into account the fact that the relic particles are disappearing. The thermal distributions satisfy the following relation (by energy conservation)
\begin{equation}
f_a^{\text{eq}} f_b^{\text{eq}} = \exp\left[- \left(E_a + E_b\right)/T\right] = \exp\left[- \left(E_c + E_d\right)/T\right] = f_c^{\text{eq}} f_d^{\text{eq}}.
\end{equation}

By using these two results and performing the integrals over the momenta, we get
\begin{equation}
\mathcal{C}_{ab \rightarrow cd} = - \langle \sigma v_{{\rm rel}}\rangle_{ab \rightarrow cd} \left[n_a n_b - \frac{n^{\text{eq}}_a n^{\text{eq}}_b}{n^{\text{eq}}_c n^{\text{eq}}_d} n_c n_d\right] = -  \langle \sigma v\rangle_{cd \rightarrow ab} \left[\frac{n^{\text{eq}}_c n^{\text{eq}}_d}{n^{\text{eq}}_a n^{\text{eq}}_b} n_a n_b - n_c n_d\right],
\label{eq:BEinv}
\end{equation}
where $\sigma$ is the total cross section for that particular process, $v_{{\rm rel}}$ is the relative velocity, and $\langle \sigma v_{{\rm rel}}\rangle_{ab \rightarrow cd}$ denotes the thermal average
\begin{equation}
\langle \sigma v\rangle_{ab \rightarrow cd} \equiv \frac{\int d^3 p_a d^3 p_b \, \sigma\left(\mathbf{p}_a,\mathbf{p}_b\right)_{ab \rightarrow cd} v_{{\rm rel}} \, f_a^{\text{eq}}(\mathbf{p}_a)f_b^{\text{eq}}(\mathbf{p}_b)}{\int d^3 p_a d^3 p_b \, f_a^{\text{eq}}(\mathbf{p}_a)f_b^{\text{eq}}(\mathbf{p}_b)}.
\label{eq:thermav}
\end{equation}

In \Eq{eq:BEinv} we have two equivalent expressions for the collision operator associated to the reaction $ab \rightarrow cd$. The first one is the usual form involving the thermal averaged cross section for that particular reaction, while the second one contains the thermal averaged cross section for the opposite reaction, $cd \rightarrow ab$.  We can use either form to write the collision operators, and we use this freedom in the $b b \chi$ model, since it is convenient to choose the reaction in which the massless degrees of freedom are in the final state in order to have a simpler expression for the thermal average.

\section{Supersymmetric QCD example}
\label{app:SUSYQCD}
In this appendix, we study a toy dark matter model based on supersymmetric QCD where the number of flavors $N_f$ and the number of colors are related by $N_f = N_c +1$ (for a model of supersymmetric QCD dark matter where $N_f < N_c$, see \Ref{Mardon:2009gw}). In the ultraviolet, the degrees of freedom are quark and antiquark fields, $Q^i$ and $\bar{Q}_{\bar{j}}$ respectively. To have a viable phenomenological model we also introduce a mass term in the superpotential
\begin{equation}
W_{{\rm UV}} = m_{i}^{\bar{j}} \, Q^i \bar{Q}_{\bar{j}}.
\label{eq:WUV}
\end{equation}
In the absence of this superpotential, the theory has a global $U(N_f)_L \times U(N_f)_R$ symmetry, which we use to put the mass matrix in a diagonal form, $m_{i}^{\bar{j}} = m_i\, \delta_{i}^{\bar{j}}$. 

Below the QCD dynamical scale, the appropriate degrees of freedom are the meson and baryon color singlet moduli, interacting via the superpotential \cite{Seiberg:1994bz}
\begin{equation}
W_{{\rm IR}} = \frac{1}{16\pi^2} \left[c^3 \, \bar{B}_j M_{ji} B_i - c^{N_f} \, \Lambda^{3 - N_f}\,{\rm det}\left(M\right) + c \,\Lambda \,{\rm Tr}\left(m M\right)\right],
\label{eq:WSUSYQCD}
\end{equation}
where we invoked ``loop-democracy'' and the $c$ factor comes from canonically normalizing the meson fields. Using naive dimensional analysis \cite{Manohar:1983md} we have $c \simeq 4\pi$. The supersymmetry preserving minimum is found by solving the system of equations
\begin{align}
\left.\frac{\partial W_{{\rm IR}}}{\partial B_i}\right|_{\Phi=\mathbf{\Phi}} & =  \frac{1}{16\pi^2} \, c^3\,\mathbf{\bar{B}}_j \mathbf{M}_{ji} = 0, \nn \\
\left.\frac{\partial W_{{\rm IR}}}{\partial \bar{B}_j}\right|_{\Phi=\mathbf{\Phi}} & = \frac{1}{16\pi^2} \,c^3\,\mathbf{M}_{ji} \mathbf{B}_i = 0, \nn \\
\left.\frac{\partial W_{{\rm IR}}}{\partial M_{ji}}\right|_{\Phi=\mathbf{\Phi}} & = \frac{1}{16\pi^2} \,\left[c^3\,\mathbf{\bar{B}}_j \mathbf{B}_i -c^{N_f} \, \Lambda^{3 - N_f}\, {\rm cof}\mathbf{M}_{ji} + c \,\Lambda \, m_{ij}\right] = 0, \label{eq:vaccum}
\end{align}
where the vacuum expectation value (vev) of a generic field $\Phi$ is denoted by boldface type $\mathbf{\Phi}$. The cofactor matrix is defined as ${\rm cof}M_{ji} = (-)^{i+j} A_{ji}$, where $A_{ji}$ is the minor of the matrix $M_{ji}$. 

The first two equations can be seen as homogeneous linear system with the baryon and antibaryons fields as unknowns and the meson matrix vev as the coefficient matrix. We assume that ${\rm det}\left(\mathbf{M}\right) \neq 0$, and then the systems in the first two lines of \Eq{eq:vaccum} have only the trivial solution $\mathbf{\bar{B}}_j = \mathbf{B}_i = 0$. Since the matrix $\mathbf{M}$ is invertible we can further simplify the vacuum condition for the meson matrix. The inverse of $M$ is given by
\begin{equation}
\mathbf{M}^{-1}_{ij} = \frac{1}{{\rm det} (\mathbf{M})}  \left[\left({\rm cof}\mathbf{M}\right)^{T}\right]_{ij} = \frac{1}{{\rm det} (\mathbf{M})}  {\rm cof}\mathbf{M}_{ji},
\end{equation}
and in conclusion the vacuum of the theory results in
\begin{equation}
\mathbf{B} = \mathbf{\bar{B}}=0, \qquad \mathbf{M} 
= c^{-1}\,\Lambda^{\frac{N_f-2}{N_f-1}} \,\left(\prod_{i=1}^{N_f} m_i\right)^{\frac{1}{N_f-1}} {\rm diag} \left(m_1^{-1}, m_2^{-1}, \ldots, m_{N_f}^{-1}\right).
\label{eq:SUSYQCDvacuum}
\end{equation}

The meson matrix vacuum expectation value is diagonal, thus we have an unbroken accidental $U(1)^{N_f}$ flavor symmetry, as well as a $U(1)_B$ symmetry acting on the baryons. The spectrum of the model is easily obtained by expanding the superpotential in \Eq{eq:WSUSYQCD} around its minimum, given in \Eq{eq:SUSYQCDvacuum}, to quadratic order:
\begin{equation}
W_{{\rm mass}} = \left[\frac{{\rm det} (m)}{\Lambda^{N_f}}\right]^{\frac{1}{N_f - 1}}
\, \frac{\Lambda^2}{m_{i}} \,\bar{B}_i B_i + \left[\frac{\Lambda^{N_f}}{{\rm det} (m)}\right]^{\frac{1}{N_f-1}} \,  \sum_{i < j} \,  \frac{m_{i}m_{j}}{\Lambda}\, \left(M_{ij} M_{ji} - M_{ii} M_{jj}\right).
\end{equation}
The baryon and the off-diagonal mesons fields are already mass eigenstates, whereas the diagonal mesons have mixing terms. In the limiting case where the mass matrix in the UV superpotential in \Eq{eq:WUV} is a multiple of the identity with all the masses equal to $m$, we have the following scaling for the masses of the IR degrees of freedom
\begin{equation}
m_B \simeq \Lambda^{\frac{N_f-2}{N_f-1}}\,m^{\frac{1}{N_f-1}}, \qquad m_M \simeq \Lambda^{\frac{1}{N_f-1}} \, m^{\frac{N_f-2}{N_f-1}}.
\label{eq:mscaling}
\end{equation}
An example spectrum is shown in \Fig{fig:SUSYQCD} for the case $N_f=4$ and for same particular values of the mass $m_i$.  In general, supersymmetry breaking effects will lead to splitting within the chiral multiplets. 

\FIGURE[t]{
\includegraphics[scale=0.5]{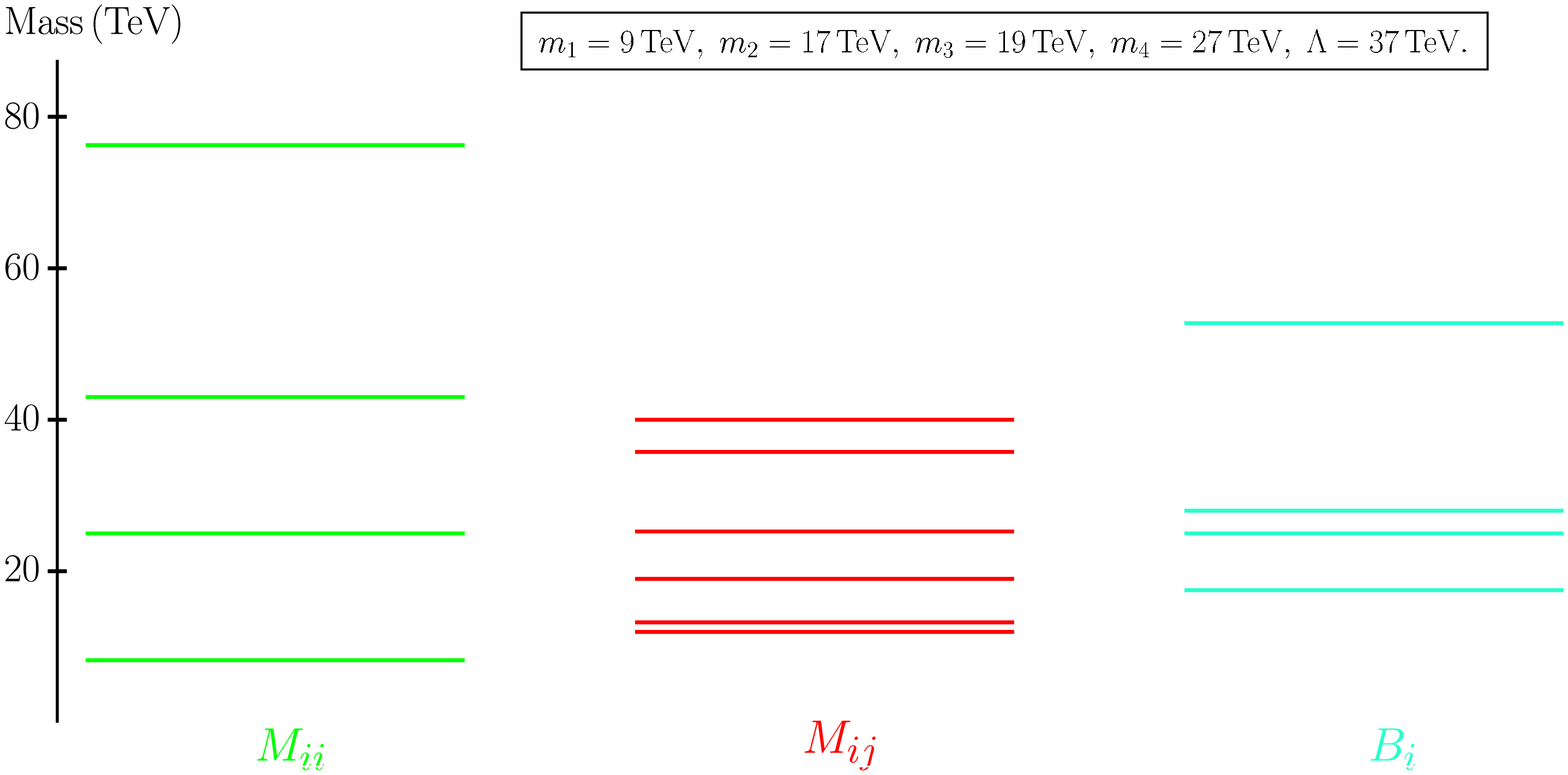}
\caption{Supersymmetric QCD spectrum for $N_f = N_c + 1 = 4$ and for particular values of the input masses $m_i$.}
\label{fig:SUSYQCD}
}

As already said, there in an unbroken $U(1)^{N_f} \times U(1)_B$ symmetry which makes the lightest baryon and the lightest off-diagonal meson absolutely stable, thus they are good dark matter candidates. The heavier baryons and off-diagonal mesons might also be stable if their decay is kinematically forbidden, as already discussed. In general a few of them can be unstable, but it is quite common to have more than just two stable components, and in the limiting case when the mass matrix $m$ is almost degenerate, all of them are stable. The diagonal mesons do not carry any flavor quantum numbers.  In analogy with the meson-baryon system in \Eq{sec:mesonbaryon} we denote them by $M_{ii} = \phi_i$, and their possible interactions with SM fields would come from the supersymmetric analogs of \Eq{eq:Vphi}.  From expanding the superpotential in \Eq{eq:WSUSYQCD}, one sees that semi-annihilations reactions involving the stable dark matter particles can take place at freeze-out, as for example
\begin{equation}
M_{12} M_{13} \rightarrow M_{23} \phi_{1}, \qquad B_2 M_{12} \rightarrow B_1 \phi_{1},
\end{equation}
where $M$, $B$, and $\phi$ can refer to either the fermions or scalars within the respective chiral multiplets.  Therefore the standard treatment for the relic density calculation cannot be applied.

\section{Thermal averages in the $bb\chi$ model}
\label{app:thaverages}
In this appendix, we derive expressions for the thermal averaged cross sections from \Eq{eq:thermav} for the $b b \chi$ model.  We work in the $s$-wave limit where the cross section does not depend on external momenta, using the amplitudes defined in \Eq{eq:bbchireacs}.

If the process $a b \rightarrow c d$ is not kinematically suppressed, then in the $s$-wave limit, the integral over momenta in \Eq{eq:thermav} is trivial and simplifies to
\begin{equation}
 \langle \sigma v\rangle_{ab \rightarrow cd} = \left[\sigma\left(\mathbf{p}_a,\mathbf{p}_b\right)_{ab \rightarrow cd} v_{{\rm rel}} 
\right]_{\mathbf{p}_a = \mathbf{p}_b = \mathbf{0}}.
\end{equation}  
It is convenient to define
\begin{equation}
w_{a b \rightarrow c d}(s) \equiv E_a E_b\, \sigma_{a b \rightarrow c d} v_{{\rm rel}} = \frac{1}{4} \int \left|\mathcal{M}_{a b \rightarrow c d}\right|^2 d\Phi^{(2)}\left(s, m_c, m_d\right),
\label{eq:w}
\end{equation}
where $s=\left(p_a+p_b\right)^2$, and $d\Phi^{(2)}\left(s, m_c, m_d\right)$ is the relativistic invariant two-body phase space. In the collision center of mass frame, we have
\begin{equation}
d\Phi^{(2)}\left(s, m_c, m_d\right) = \frac{d \Omega}{32\pi^2} \left[1 + \frac{(m_c^2-m_d^2)^2 - 2 s (m_c^2+m_d^2)}{s^2}\right]^{1/2},
\end{equation}
where $d \Omega$ is the differential solid angle element.  Plugging this expression into \Eq{eq:w}, and assuming that the matrix element has no angular dependence (as is the case in the $s$-wave limit), we find
\begin{equation}
w_{a b \rightarrow c d}(s) = \frac{\left|\mathcal{M}_{a b \rightarrow c d}\right|^2}{32\pi} \left[1 + \frac{(m_c^2-m_d^2)^2 - 2 s (m_c^2+m_d^2)}{s^2}\right]^{1/2}.
\end{equation}
Thus, the thermally averaged cross section is
\begin{equation}
 \langle \sigma v\rangle_{ab \rightarrow cd} = \left[\frac{w_{a b \rightarrow c d}(s)}{E_a E_b}\right]_{\mathbf{p}_a = \mathbf{p}_b = \mathbf{0}} = \frac{w_{a b \rightarrow c d}\left( (m_a +m_b)^2 \right)}{m_a m_b}.
 \label{eq:easythermal}
\end{equation}  

We can use \Eq{eq:easythermal} to compute the thermal averages for all the reactions in the $b b \chi$ model, except for the reactions $b \bar{b} \leftrightarrow \chi\bar{\chi}$ where the presence of a kinematic threshold requires more care \cite{Griest:1990kh}. For future reference we also define $r=m_b / m_\chi$, and we give all the results in terms of $m_{\chi}$ and $r$. We have three different reaction types: annihilation, semi-annihilation, and dark matter conversions.

\subsubsection*{Annihilations}
The ordinary annihilation reactions are $b \bar{b} \rightarrow \phi\phi$ and $\chi \bar{\chi}  \rightarrow \phi\phi$. In this case, we have two effectively massless final states, thus
\begin{equation}
\langle \sigma v_{{\rm rel}}\rangle_{b \bar{b} \rightarrow \phi\phi} = \frac{1}{4 r^2} \frac{\alpha^2}{32\pi m_{\chi}^2},
\end{equation}
where the factor of $1/4$ comes from the average over the spin degrees of freedom, and
\begin{equation}
\langle \sigma v_{{\rm rel}}\rangle_{\chi \bar{\chi}  \rightarrow \phi\phi} = \frac{\beta^2}{32\pi m_\chi^2}.
\end{equation}
 
\subsubsection*{Semi-annihilations}
In the semi-annihilation case, we have only one massless particle in the final state, thus there is an additional phase space suppression
\begin{equation}
\langle \sigma v_{{\rm rel}}\rangle_{b b \rightarrow \phi\bar{\chi}} = \frac{1}{4 r^2} \left(1 - \frac{1}{4 r^2}\right) \frac{\epsilon^2}{32\pi m_\chi^2},
\end{equation}
and
\begin{equation}
\langle \sigma v_{{\rm rel}}\rangle_{\chi b \rightarrow \phi\bar{b}} = \frac{1}{2 r} \left(1 - \frac{r^2}{(1+r)^2}\right) \frac{\epsilon^2}{32\pi m_\chi^2}. 
\end{equation}

\subsubsection*{Dark matter conversions}
In this last case, we have two massive particles in the final state. The phase space now is suppressed, and the reaction can be forbidden at zero relative velocity. To correctly compute the thermal average we consider the second exception discussed in \Ref{Griest:1990kh}, namely when the reaction takes place near a kinematic threshold. 

We start from the reaction $b \bar{b} \rightarrow \chi\bar{\chi}$. The Mandelstam variable in the center of mass frame of the collision is
\begin{equation}
s=\left(E_{\chi}+E_{\bar{\chi}}\right)^2 = 4 m_\chi^2 \gamma^2, \qquad \qquad \left[1 - \frac{4 m_\chi^2}{s}\right]^{1/2} = \left[1 - \frac{1}{\gamma^2}\right]^{1/2} = v_2,
\end{equation}
where $v_2$ is the velocity of the $\chi$ particles in the final state in the center of mass frame for the collision. In the $s$-wave limit we can write
\begin{equation}
w_{b \bar{b} \rightarrow \chi\bar{\chi}}(s) = \frac{1}{4} \frac{\kappa^2}{32\pi}\,v_2 ,
\end{equation}
where the factor of $1/4$ is because the average over the spin degrees of freedom. We define
\begin{equation}
\mu_{b} = \left|r^2 - 1\right|^{1/2},
\end{equation}
and for the thermal average we have the following expression \cite{Griest:1990kh}
\begin{equation}
\langle \sigma v_{{\rm rel}}\rangle_{b \bar{b} \rightarrow \chi\bar{\chi}} = \frac{1}{4 r^2} \frac{\kappa^2}{32\pi m_\chi^2}\, \frac{\mu_{b}^2  \, x^{1/2}}{r^{1/2} \pi^{1/2}} \, e^{\pm \mu_{b}^2 r x /2}\, K_1\left[\mu_{b}^2 r x /2\right],
\end{equation}
where the $+$ sign ($-$ sign) is for the case when the annihilation if allowed (forbidden) at zero relative velocity in the initial state.

For the opposite reaction, $\chi\bar{\chi} \rightarrow b \bar{b}$, the discussion is analogous, up to some factor. We do not have to average over any initial spin, and we can write
\begin{equation}
w_{\chi\bar{\chi} \rightarrow b \bar{b}}(s) = \frac{\kappa^2}{32\pi}\,v_2 ,
\end{equation}
where $v_2$ is the velocity of the $b$ particles in the final state in the center of mass frame for the collision. We define
\begin{equation}
\mu_{\chi} = \left|\frac{1}{r^2} - 1\right|^{1/2},
\end{equation}
and for the thermal average we apply again the results of \Ref{Griest:1990kh}
\begin{equation}
\langle \sigma v_{{\rm rel}}\rangle_{\chi\bar{\chi} \rightarrow b \bar{b}} = \frac{\kappa^2}{32\pi m_\chi^2}\, \frac{\mu_{\chi}^2 \,r \, x^{1/2}}{\pi^{1/2}} \, e^{\pm \mu_{\chi}^2 x /2}\, K_1\left[\mu_{\chi}^2 x /2\right],
\end{equation}
where $\pm$ for the allowed/forbidden reaction at zero relative velocity in the initial state, respectively.


\bibliographystyle{JHEP}
\bibliography{FourthException}

\end{document}